\documentclass[aps,pra,reprint,twocolumn,superscriptaddress,showpacs,showkeys,superscriptaddress,longbibliography]{revtex4-1}

\usepackage{amsmath,amssymb,amstext}
\usepackage[usenames,dvipsnames]{color}
\usepackage{graphicx}
\usepackage{bm,bbold,braket}
\usepackage{natbib}
\usepackage{txfonts, comment,stmaryrd}
\usepackage{dcolumn}

\usepackage[english]{babel}

\usepackage[colorlinks,bookmarks=false,citecolor=blue,linkcolor=red,urlcolor=blue]{hyperref}
\begin{document}

\title{Dynamics of Rydberg excitations and quantum correlations in an atomic array coupled to a photonic crystal waveguide}
\author{Yashwant Chougale}
\email{yashwant.chougale@students.iiserpune.ac.in}
\affiliation{Indian Institute of Science Education and Research, Pune 411 008, India}
\author{Jugal Talukdar}
\email{jugal.talukdar@ou.edu}
\affiliation{Indian Institute of Science Education and Research, Pune 411 008, India}    
\affiliation{Center for Quantum Research and Technology, The University of Oklahoma, Norman, Oklahoma 73019, USA}
\affiliation{Homer L. Dodge Department of Physics and Astronomy, The University of Oklahoma, 440 W. Brooks Street, Norman, Oklahoma 73019, USA}
\author{Tom\'as Ramos}
\email{t.ramos.delrio@gmail.com}
\affiliation{Instituto de F\'isica Fundamental IFF-CSIC, Calle Serrano 113b, 28006 Madrid, Spain}
\affiliation{DAiTA Lab, Facultad de Estudios Interdisciplinarios, Universidad Mayor, Santiago, Chile}

\author{Rejish Nath}
\email{rejish@iiserpune.ac.in} 
\affiliation{Indian Institute of Science Education and Research, Pune 411 008, India}

\begin{abstract}
We study the dynamics of up to two Rydberg excitations and the correlation growth in a chain of atoms coupled to a photonic crystal waveguide. In this setup, an excitation can hop from one atom to another via exponentially decaying exchange interactions mediated  by  the  waveguide. An initially localized excitation undergoes a continuous-time quantum walk for short-range hopping, and for long-range, it experiences quasi-localization. Besides that, the inverse participation ratio reveals a super-ballistic diffusion of the excitation in short times, whereas, at a long time, it becomes ballistic. For two initially localized excitations, intriguing, and complex dynamical scenarios emerge for different initial separations due to the competition between the Rydberg-Rydberg and exchange interactions. In particular, the two-point correlation reveals a light-cone behavior even for sufficiently long-range exchange interactions. Additionally, we characterize the growth of bipartite entanglement entropy, which exhibits a global bound if only one excitation is present in the dynamics. Finally, we analyze the effect of imperfections due to spontaneous emission from the Rydberg state into photons outside the waveguide and show that all  physical phenomena we predict are well within experimental reach.
\end{abstract}

\pacs{}

\keywords{}

\maketitle

\section{Introduction}

Engineering strong atom-photon interactions \cite{joh90,she05,kie05,dzs10,kie08,cha07,sha13} has become the holy grail in most quantum optical systems including ultra-cold atoms \cite{cha12,has10,gul12,chan13,dou15,vet10,gob12} due to their potential applications in quantum information protocols \cite{kim08,cir97,dua04}, as well as in the exploration of novel quantum many-body physics \cite{har06,gree06,ang07,gop09,bau10,dou15,hun16,manz17}. Recent focus has been on integrating nanophotonics with atomic physics via photonic crystal waveguides (PCWs) \cite{gob14, hoo16, yu14,hun13,manz17, sol17,beg19,cor19}. PCWs are periodic dielectric structures exhibiting photonic band gaps, which can be used to control and modify the light propagation. The location and the size of the band gaps can be adjusted, for instance, by changing the periodic dielectric function of the PCW \cite{pcb}. Atoms trapped near a PCW act as dielectric defects, which develop localized atom-photon bound states in the bandgap regime \cite{joh90,sha13, tho13,dou15, kur90, john91, gon15}. Effectively, this leads to long-range atom-atom interactions mediated by waveguide photons and allow the simulation of exotic spin models with a high degree of controllability over the range and the nature of the spin-spin interactions \cite{dou15, gon15, hun16, liu16}. 

In this paper, we look at the dynamics of Rydberg-excitations in a chain of single atoms coupled to a PCW. Similar setups include Rydberg excitations in a hollow-core photonic crystal fiber \cite{epp14,lan17}, and an optical nanofiber \cite{kri19}. The excitations are exchanged to other ground-state atoms via locally induced cavity modes.  These exchange or "hopping" interactions decay exponentially over distance, with a high tunability over the range of interactions \cite{dou15}. An initially localized single excitation undergoes quantum diffusion, and the properties of the excitation dynamics depend crucially on the range of the exchange potential. For instance, for short-range exchange couplings, the long-time probability distribution (LTPD) for the position of an excitation exhibits the features of a continuous-time quantum walk (CTQW) \cite{far98,pat05,man07,per08,oli11,ven12,por15}. In contrast, for long-range, LTPD displays quasi-localization and tailing behavior. The quasi-localization is due to the flat modes near the edges of the Brillouin zone, and the tailing behavior emerges from the long-wavelength modes.  A better understanding of the excitation dynamics is accessed through excess kurtosis and inverse participation ratio (IPR). Interestingly, IPR reveals to us that the single excitation dynamics is super-ballistic at the very initial stage of the quantum diffusion due to the initially localized state of the excitation, whereas, at longer times, it is ballistic. The initially localized excitation being a source of entangled pairs of quasi-particles with opposite momenta correlates different parts of the atomic array during the time evolution. We quantify the correlation via the bipartite entanglement entropy for which the system is divided into two equal parts. Interestingly, we found that the entanglement entropy generated by a single excitation is globally bounded, independently of the system size. The single excitation scenario we discuss here can be related to an impurity in a quantum spin chain, with its relevance in entanglement or quantum state transfer \cite{zwi11}. On the other hand, the photon transport is studied using an identical setup \cite{son18,son19}.  

%%%%%%%%%%%%%%
\begin{figure}[hbt]
\vspace{0.cm}
\centering
\includegraphics[width= 1.\columnwidth]{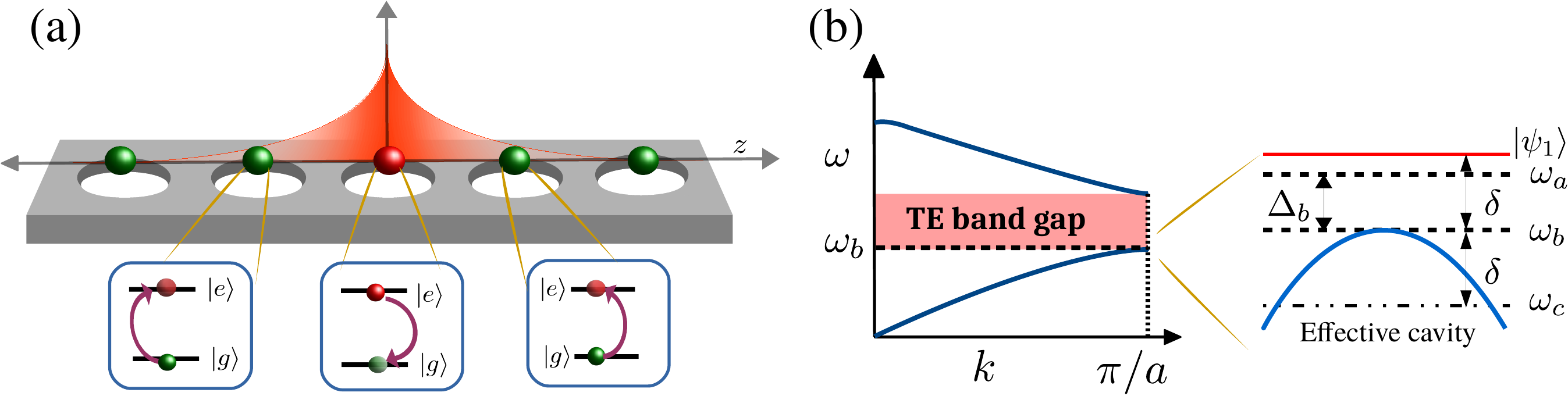}
\caption{\small{(Color online) (a) Atomic lattice and PC setup. An exponentially decaying single-photon profile is depicted at the position of the central atom and mediated by it, the excitation can be exchanged to other ground state atoms. (b) Band structure of a PC, with the atom-band edge detuning $\Delta_b=\omega_a-\omega_b$ where $\omega_a$ is the atomic resonance frequency, and $\omega_b$ provides the lowest band edge. $\omega_c$ is the effective cavity mode frequency with an effective detuning $\delta$. The eigenfrequency $\omega$ of the atom-photon bound state $|\psi_1\rangle$ lies in the bandgap.}}
\label{fig:1} 
\end{figure}

In the presence of two excitations, the Rydberg-Rydberg interactions (RRIs) play an essential role in the dynamics. To separate the effects of long-range hopping and RRIs, we first analyze the dynamics of two non-interacting excitations, observing spatial anti-bunching for short-range hoppings and particle quasi-localization for long-range. For intermediate-range, one of the excitations gets quasi-localized, and the other propagate away from it. On the other hand, when including the RRIs, we observe the emergence of bound-states in the excitation spectrum above a critical interaction strength. Besides the well-known dynamical features found for fermions or hardcore bosons in an optical lattice \cite{qin14, beg18, lah12, fuk13, pre15,  ben12, cha16, val08, win06, pii07, pet07, pii07, val09}, we predict new features in our system due to the competition between the RRIs and the long-range exchange couplings. For instance, above a critical interaction strength, the initially localized excitations may carry both bound and scattering quasi-particles leading to intricate patterns in the probability distributions at longer times. In contrast to the single excitation case, we found that the bipartite entanglement of two excitations is not globally bounded but depends on the initial separation between them as well as the range of the exchange couplings. Also, the two-point correlations reveal a light-cone behavior even for sufficiently long-range exchange interactions. Finally, we look at the effect of decoherence in a realistic implementation of our setup, such as the spontaneous emission from the Rydberg state. We show that all the features of the coherent dynamics survive up to reasonable decay rates and therefore the observation of these intricate Rydberg dynamics is within experimental reach.

The paper is structured as follows. In Sec \ref{sp}, we discuss the setup and the Hamiltonian describing the system. The coherent dynamics of the single excitation is analyzed in Sec. \ref{sed}, including the growth of bipartite entanglement due to the delocalized excitation. In Sec. \ref{2ed}, we discuss the dynamics of two excitations and the correlations. Finally, in Sec. \ref{diss}, we briefly discuss the dissipative dynamics of the system, incorporating the spontaneous emission rate of the atomic excitation. 

%%%%%%%%%%%%%%%%
\section{Setup and Model}
\label{sp}
Our work is motivated by the recent developments in coupling atoms to a PC, and in particular, based on the setup described in Ref.~\cite{dou15}. Consider a one-dimensional setup, which consists of an array of $N$ two-level atoms with a ground state $|g\rangle$ and an excited Rydberg state $|e\rangle$, trapped near a PC and arranged along waveguide longitudinal $z$-axis as shown in Fig.~\ref{fig:1}(a). We assume deep potential wells so that the center of mass motion of the atoms can be neglected, and they remain in their respective lattice sites $j=1,...,N$. The atomic resonant frequency $\omega_a$ lies in the bandgap of the PC, with a detuning $\Delta_b=\omega_a-\omega_b$ from the lowest band edge $\omega_b$ [Fig.~\ref{fig:1}(b)]. Other band edges are assumed very far from $\omega_a$, and therefore, only the modes close to $\omega_b$ play an important role. The dispersion can then be approximated as $\omega_k\approx\omega_b[1-\alpha(k-k_0)^2/k_0^2]$, where $k_0$ is the band edge wave number, and $\alpha$ determines the band curvature. These modes are of Bloch form due to the periodic structure of the PC, with electric field $E_{k_0}(z)=e^{ik_0z}u_{k_0}(z)$. The Hamiltonian that describes a single two-level atom trapped at $z=0$ coupled to the PCW reads, 
\begin{equation}
\hat H=\hbar \omega_a\hat \sigma_{ee}+\hbar\int dk \omega_k \hat a_k^{\dagger}\hat a_k+\hbar g\int dk(\hat \sigma_{eg}\hat a_kE_k(0)+h.c.),
\end{equation}
where the atomic operators $\hat\sigma_{\alpha\beta}=|\alpha\rangle\langle \beta|$ with $\alpha, \beta\in \{e, g\}$ and $\hat a_k$ ($\hat a_k^{\dagger}$) is an annihilation (creation) operator of a photon in the $k^{th}$ mode. The atom-light coupling constant is $g=d_{eg}\sqrt{\omega_b/4\pi\hbar\epsilon_0A}$ with $d_{eg}$ being the atomic dipole moment, and $A$ is the mode cross-sectional area \cite{hun13}. In the single excitation sector of the system, there exists a bound eigenstate ($\hat H|\psi_1\rangle=\hbar\omega|\psi_1\rangle$) in which the atom is dressed by a localized excitation of the photonic mode. The eigenfrequency  $\omega$ lies at the bandgap with a detuning $\delta=\omega-\omega_b$, and it has been shown that the eigenenergy can be approximately obtained from the positive root of  $(\delta-\Delta)\sqrt{\delta}=\beta^{3/2}$ with $\beta=(\pi g^2|u_{k_0}(0)|^2k_0/\sqrt{4\alpha\omega_b})^{2/3}$ \cite{dou15}. The single excitation bound state is of the form: $|\psi_1\rangle=\cos\theta |e\rangle |0\rangle+\sin\theta |g\rangle|1\rangle$, where the atom is dressed by a localized photon: $|1\rangle=\int dk c_k \hat a_k^{\dagger}|0\rangle$ around the atomic position and has the spatial wave function 
\begin{equation}
\phi(z)=\sqrt{\frac{2\pi}{L}}e^{-|z|/L}E_{k_0}(z).
\end{equation}
The length scale,  $L=\sqrt{\alpha\omega_b/(k_0^2\delta)}$ quantifies the exponential decay of the photon probability from the atomic position. Interestingly, one can map this scenario to that of an atom-cavity system with a cavity length $L$, cavity mode frequency $\omega_c=\omega_b-\delta$, effective atom-cavity coupling $g_c=g\sqrt{2\pi/L}$,  and an effective detuning $\Delta_c=\Delta_b+\delta$ [see Fig.~\ref{fig:1}(b)]. With all these, the state $|\psi_1\rangle$ is mapped into the corresponding dressed state in the Jaynes-Cummings model. Now extending to an array of atoms with positions $z_j$, and assuming the far-detuned limit: $\Delta_c\gg \beta$, the weakly populated photonic modes can be adiabatically eliminated. This leads to a dipole-dipole exchange Hamiltonian for the atomic excitations of the form, 
\begin{equation}
\hat H_{ex}=J\sum_{j,l}^N\hat\sigma_{eg}^j\hat\sigma_{ge}^lf(z_j,z_l),
\label{h1}
\end{equation}
where the exchange coupling strength reads $J=\hbar g_c^2/\Delta_c$. We take $J$ as a constant and explore the effect of $L$ on the excitation dynamics, which appears in the exponential function, $f(z_j, z_l)=\exp(-|z_j-z_l|/L)E^*_{k_0}(z_j)E_{k_0}(z_l)$. Taking lattice spacing as twice the length of the PC unit cell, we have $E_{k_0}(z_j)=1$ at atomic locations \cite{dou16}. As we can see that the range of the exchange interactions, $L$, can be controlled by tuning the band structure of the PC. 

In the following, we analyze the coherent dynamics induced exclusively by the spin-spin Hamiltonian in Eq.~\ref{h1}, for both one and two excitations. Then, we include RRIs between the atoms and finally, extend the analysis to the dissipative case by including imperfections due to unwanted dissipation of the excited levels into photons outside the waveguide.

%%%%%%%%%%%%%%%%%%%%%%
\section{One Excitation}
\label{sed}
\begin{figure}
\vspace{0.cm}
\centering
\includegraphics[width= 1.\columnwidth]{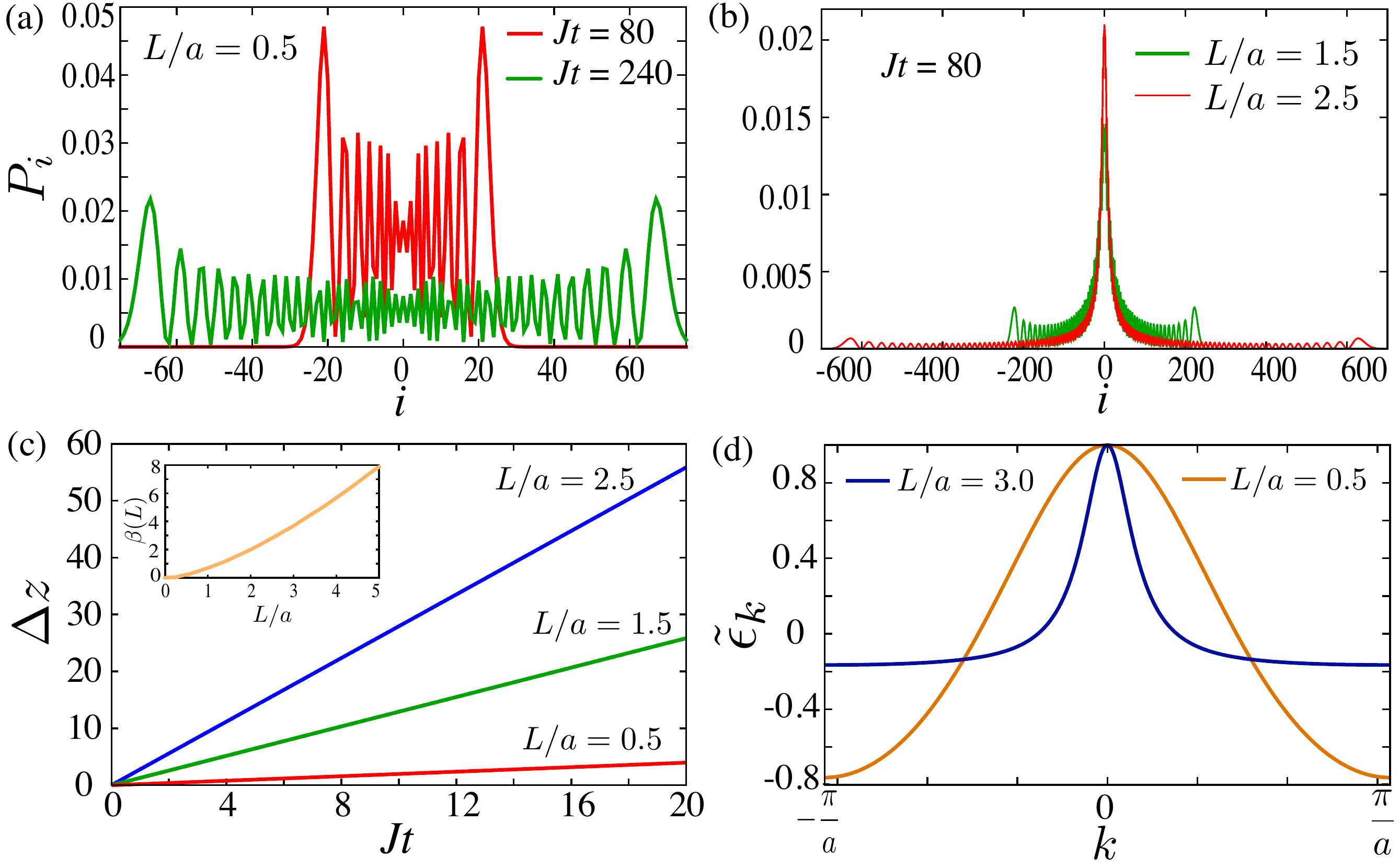}
\caption{\small{(Color online) (a) The probability density $P_i$ as a function of lattice index $i$ for $L/a=0.5$ at two different $t$. At $t=0$, the excitation is initially localized at $i=0$. $P_i$ exhibits a typical profile of CTQW at longer times. (b) $P_i$ vs $i$ for different $L$ at a given instant of time $Jt=80$. (c) The spread $\Delta z$ of the distribution as a function of time for different $L$ values. The inset shows $\beta(L)$, the rate at which the spread of $P_i$ increases in time, as a function of $L$. (d) The spectrum $\tilde\epsilon_k=\epsilon_k/|\epsilon_{k=0}|$ vs $k$ in the first Brillouin zone for different $L$. For large $L$, the energy band becomes increasingly flat near the edges of the Brillouin zone and steeper at very low momenta, giving rise to quasi-localization and tailing behavior in LTPD.}}
\label{fig:2} 
\end{figure}   
%%%%%%
The single excitation dynamics is studied by numerically solving the Schr\"odinger equation: $i\partial|\psi(t)\rangle/\partial t=H_{ex}|\psi(t)\rangle$ ($\hbar=1$). We assume the excitation is initially localized at the center of the lattice, and evolve the system up to various times before the excitation probability hits the boundary. The Hamiltonian $H_{ex}$ preserves the number of excitations, and we can truncate the Hilbert space to the subspace of $N$ singly excited states $\{|i\rangle\equiv |...g^{i-1} e^i g^{i+1}... \rangle \}$. The probability of finding an excitation at the site $i$ is given by $P_i(t)=|\langle i|\psi(t)\rangle|^2$. Fig. \ref{fig:2}(a) shows $P_i$ vs $i$ for $L/a=0.5$, which exhibits a probability distribution typical to that of a CTQW, with the maximum value of $P_i$ at the edges of the distribution at long time. The spread of the distribution $\Delta z=\sqrt{\langle z^2\rangle-\langle z\rangle^2}$ increases linearly in time [see Fig. \ref{fig:2}(c)], where $\langle z^n\rangle=\langle\psi(t)|z^n|\psi(t)\rangle$. The linear behavior of $\Delta z$ in $t$ is attributed to quantum interference and is in high contrast to the case of a classical random walk. The latter case is characterized by a Gaussian distribution at large times, with $\Delta z\propto\sqrt{t}$. 
%%%%%

\subsection{Quasi-localization and Tailing}
Interestingly, as $L$ increases, especially when $L>a$ the LTPD gets modified drastically as shown in Fig.~\ref{fig:2}(b). It exhibits a quasi-localization behavior, possessing a sharp peak at the center of the lattice with a long tail. At a given instant, an increment in $L$ makes the central peak sharper with higher values, and a longer tail [see Fig. \ref{fig:2}(b)]. Despite having a qualitative change in the shape of LTPD for large $L$, its spread $\Delta z$ remains a linear function of time with an $L$-dependent proportionality constant, i.e., $\Delta z=\beta(L) t$. The latter implies that the diffusion of the initially localized excitation is {\em ballistic} for any value of $L$. The rate of quantum diffusion parameter $\beta(L)$ increases monotonously with $L$ as shown in the inset of Fig. \ref{fig:2}(c). In the large $N$ limit, we get an analytic expression for $\beta(L)$ using the relation \cite{ale12}: 
\begin{equation}
\langle z^2\rangle(t)=\left[ \frac{a}{2\pi} \int_{BZ} \left(\frac{d\epsilon_k}{dk}\right)^2 dk\right] t^2
\label{m2}
\end{equation}
where $BZ$ stands for the first Brillouin zone. The free particle dispersion $\epsilon_k$ of the Hamiltonian $\hat H_{ex}$ [Eq. (\ref{h1})] is obtained as
\begin{equation}
\epsilon_k=J\frac{\left(\cos ka-e^{-a/L}\right)}{\cosh (a/L)-\cos ka},
\label{spe}
\end{equation}
having a bandwidth of $2J/\sinh(a/L)$. Using Eq. (\ref{spe}) in Eq. (\ref{m2}), we get
\begin{equation}
\beta(L)=aJ\sqrt{\frac{\coth(a/L)}{2\sinh^2(a/L)}},
\label{alp}
\end{equation}
that is in an excellent agreement with the numerical results for sufficiently large $N$.
%%%%%%%%

The quasi-localization and the tailing behavior of LTPD at large $L$ can be understood using the energy spectrum $\epsilon_k$. As $L$ increases, near the edges of the Brillouin zone, the spectrum gets increasingly flat  [see Fig. \ref{fig:2}(d)], resulting in a vanishingly small group velocity $v_g(k)=d\epsilon_k/dk$ for modes with high momenta, whereas for low momenta the group velocity $v_g$ becomes increasingly large. This implies that the quasi-localization is due to the flat modes near the edges of the Brillouin zone, whereas the fast propagating tail is arising from the long-wavelength modes.

\subsection {Kurtosis and Inverse participation ratio}
To gain a comprehensive picture of the excitation dynamics, and to characterize the behavior of the LTPD as a function of $L$, we look at both the excess kurtosis of the probability distribution and the inverse participation ratio (IPR). 

%%%%
\begin{figure}
\vspace{0.cm}
\centering
\includegraphics[width= 1.\columnwidth]{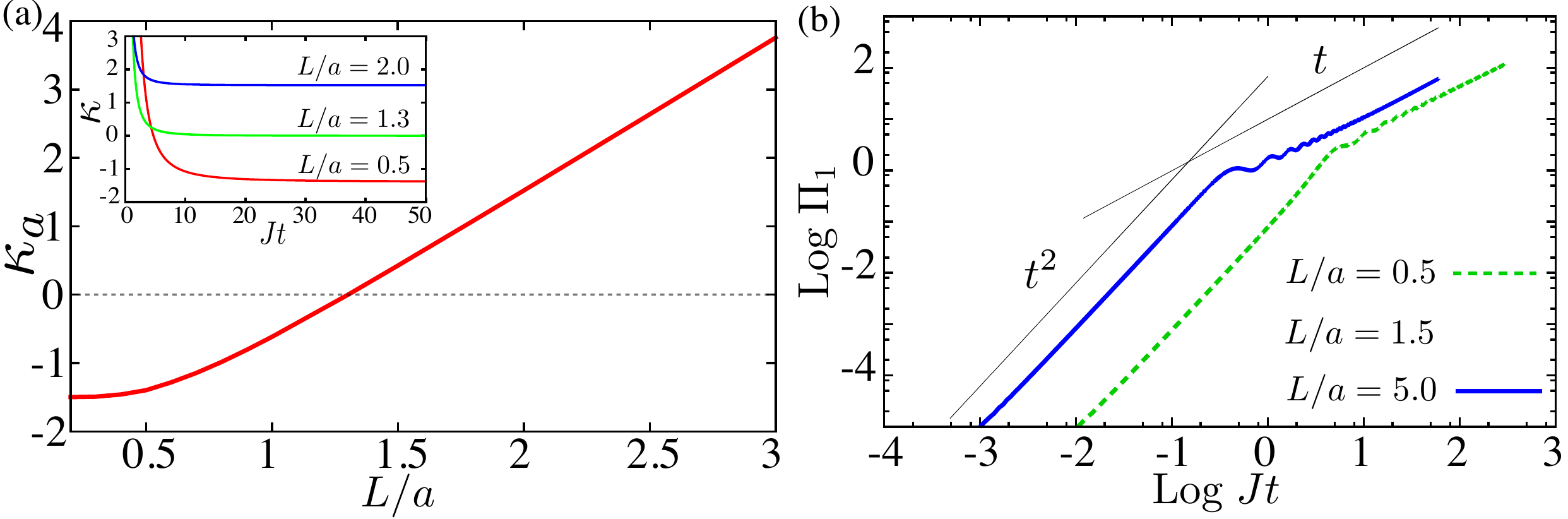}
\caption{(Color online) \small{(a) The numerical results for the asymptotic value of the excess kurtosis ($\kappa_a$) vs $L$ for $N=2001$. The inset of the (a) shows the time dependence of the kurtosis for different values of $L$. The positive $\kappa_a$ for $L/a>1.33$ characterizes the quasi-localization, and negative $\kappa_a$ implies CTQW dynamics. (b) IPR $[\Pi_1(t)]$ vs time for different $L$. The solid thin lines symbolically show the $t^2$ and $t$ behavior, indicating the super-ballistic at short times and ballistic diffusion at longer times, respectively.}}
\label{fig:3} 
\end{figure}  
{\em Kurtosis}:---In probability theory, the peakedness and tailing of a probability distribution are characterized in terms of a dimensionless parameter called excess kurtosis ($\kappa$), which is defined as, 
\begin{equation}
 \kappa=\frac{\langle(z-\langle z\rangle)^4\rangle}{\langle(z-\langle z\rangle)^2\rangle^2}-3.
 \label{kur}
\end{equation}
The first term in Eq. (\ref{kur}) is the normalized $4$-th central moment of the probability distribution. A positive-valued $\kappa$ implies a distribution that is more peaked around the mean value with a lengthy tail, as compared to a Gaussian distribution for which $\kappa=0$. Similarly, $\kappa<0$ corresponds to distributions having a flat head with a fast decaying tail. In the context of single excitation dynamics discussed above, a significantly large and positive $\kappa$ implies a (quasi) localized state, whereas a negative $\kappa$ signifies a typical CTQW profile.  The inset of Fig. \ref{fig:3}(a) shows the time dependence of $\kappa$ for different values of $L$. $\kappa$ decreases in time, and asymptotically ($t\to \infty$) approaches a constant value ($\kappa_a$). A constant $\kappa_a$ at longer times indicates that the envelope of the probability distribution is invariant in time, even though it spreads. The quantity $\kappa_a$ exhibits a smooth cross over (around $L/a\sim 1.3$) from a negative to a positive value as a function of $L$ [see Fig. \ref{fig:3}(a)] and for large $L$, it increases linearly with $L$. The latter indicates that both the peakedness and the tailing behavior get augmented at large $L$. An analytic expression for the kurtosis is obtained in the large $N$ limit using the relation,
\begin{equation}
\langle z^4\rangle(t)=\frac{a}{2\pi} \int_{BZ} \left[t^2\left(\frac{d^2\epsilon_k}{dk^2}\right)^2+t^4\left(\frac{d\epsilon_k}{dk}\right)^4 \right] dk,
\label{m4}
\end{equation}
and we get,
\begin{equation}
\kappa(t)=\frac{2[2+\cosh^2(a/L)]\tanh(a/L)}{(Jt)^2}+\frac{3[1+2\cosh^2(a/L)]}{\sinh(2a/L)}-3,
\end{equation}
which gives us, $\kappa_a=3[1+2\cosh^2(1/L)]/\sinh(2/L)-3$ and it agrees exactly with the numerical results for sufficiently large $N$. For large $L$, the asymptotic kurtosis, $\kappa_a\approx 9L/(4a)-3$, depends linearly on $L$ as expected. The derivation for the analytical results for the kurtosis and the eigenspectrum are given in appendix \ref{a1}.
 
%%%%%% 

{\em IPR}:-- Interestingly, the inverse participation ratio (IPR), 	
\begin{equation}
\Pi_1(t) = \frac{1}{\sum_{i}P_i^2}-1 
\end{equation}
captures some fine details of the dynamics, especially at the initial stages of the diffusion.  For a completely localized excitation, we have $\Pi_1= 0$ and for a completely delocalized case we get the uniform distribution, $\Pi_1=N-1$, which suggests that $\Pi_1$ can be interpreted as a sort of length over which the excitation is delocalized. As previously discussed, the linear dependence of the width of the probability distribution with time ($\Delta z\propto t$) indicates that the diffusion is ballistic at any instant of time. Nevertheless, IPR predicts that at the very initial stage of the time evolution, i.e., for times $Jt\leq 1$, we have $\Pi_1(t)\propto t^2$, indicating a super ballistic diffusion [see Fig. \ref{fig:3}(b)]. At larger times, $\Pi_1(t)\propto t$, the diffusion is ballistic. The transient super-ballistic nature of IPR is due to the initially localized state of the excitation  \cite{ste15}. Further, at a given instant of time, IPR is larger for larger $L$ in the super ballistic regime, but in the ballistic regime $\Pi_1(t)$ first increases with $L$ and then decreases due to the quasi-localization at sufficiently large $L$. Thus, Kurtosis captures the structural change in the LTPD as a function of $L$ whereas IPR gives us insights into the diffusive nature of the excitation.
  %%%%%%%%
   
 \subsection{Bipartite Entanglement entropy}
 \label{seee}

 %%%%%
\begin{figure}
\centering
\includegraphics[width= .9\columnwidth]{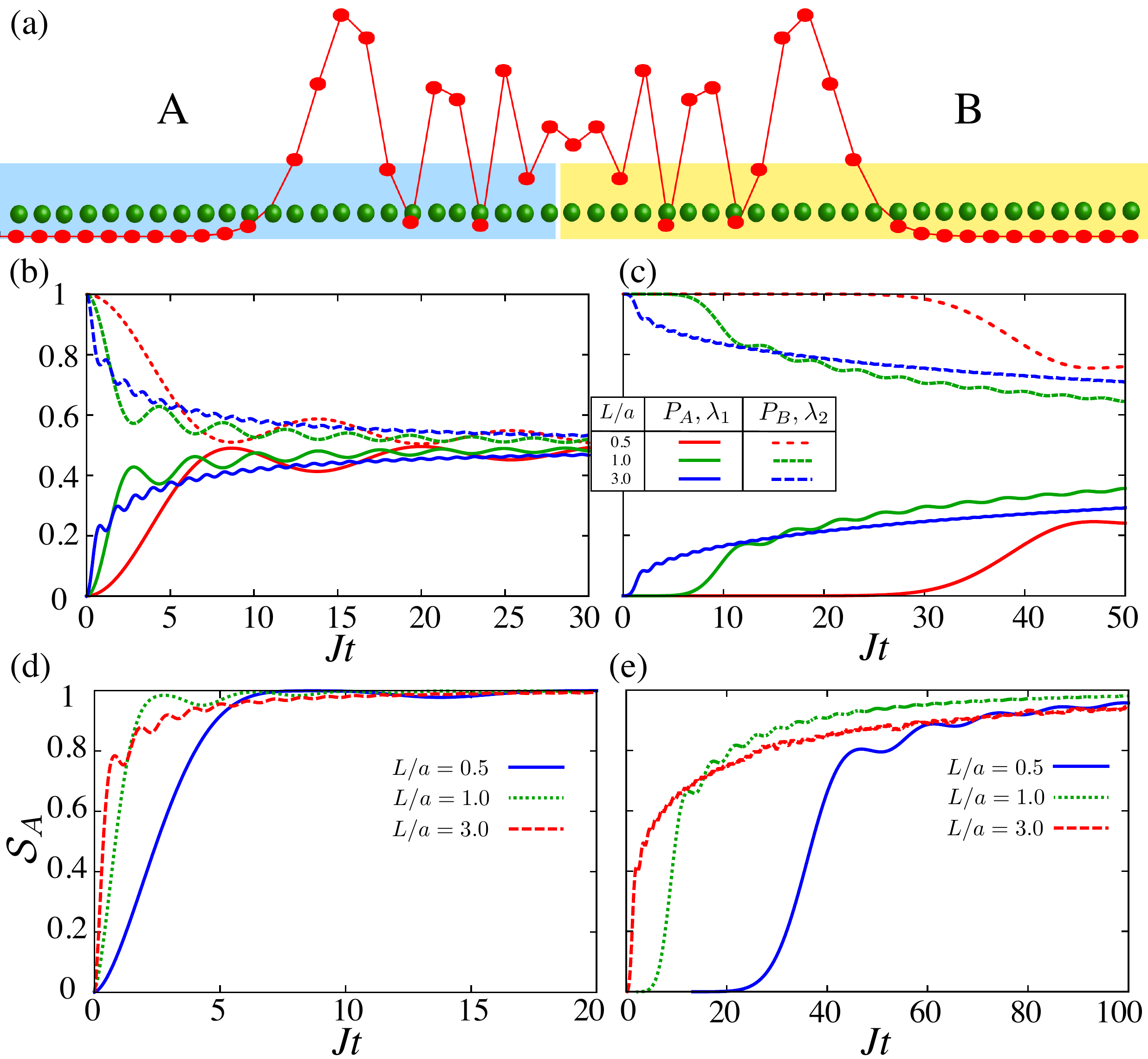}
\caption{\small{(a) The partition of the system into two identical subsystems ($A$ and $B$) and the excitation is delocalized over the atomic array. (b) and (c) show the eigenvalues of the density operator $\rho_A$ which provide us the probability of finding the excitation in the subsystems $A$ and $B$ where initially the excitation is placed in the subsystem $B$. For (b) $|\psi(t=0)\rangle=|0\rangle_A|N/2+1\rangle_B$ and for (c) $|\psi(t=0)\rangle=|0\rangle_A|N/2+10\rangle_B$. The growth of entanglement entropy $\mathcal S_A$ in time for different $L$ for the parameters in (b) and (c) are respectively shown in (d) and (e).}}
\label{fig:4} 
\end{figure}
 
 As we have seen, the excitation gets delocalized over the lattice under the unitary evolution governed by the Hamiltonian in Eq. (\ref{h1}). It implies that the different parts of the lattice get correlated via the delocalized excitation \cite{jur14}. To quantify this effect, we divide the atomic array into two parts $A$ and $B$ in which $A$ constitutes the left half, and $B$ represents the right half with the number of lattice sites $N/2$ each. The Hilbert spaces for the subsystems, each having a dimension of $N/2+1$, are spanned by $\mathcal H_A\in \{|0\rangle_A, |i\rangle_A\}$ for $A$ and $\mathcal H_B\in \{|0\rangle_B, |i\rangle_B\}$ for $B$. The state $|0\rangle_A$ ($|0\rangle_B$) represents all $N/2$ atoms in the subsystem $A$ ($B$) are in the ground state $|g\rangle$ and  $|i\rangle_A$ ($|i\rangle_B$) is the singly excited state in which only the atom in the $i$th site is in the state $|e\rangle$. As far as the excitation is concerned, each subsystem can be visualized as an effective two-level system with two states that stands for the presence and absence of the excitation. Now, we can write $|\psi(t)\rangle=\sum_{i=1}^{N/2}c_i(t)|i\rangle_A|0\rangle_B+\sum_{i=N/2+1}^{N}c_i(t)|0\rangle_A|i\rangle_B$, and the density matrix of the total system is $\rho_{AB}=|\psi(t)\rangle\langle \psi(t)|$. Thus, $P_A(t)=\sum_{i=1}^{N/2}|c_i(t)|^2$ and $P_B(t)=\sum_{i=N/2+1}^{N}|c_i(t)|^2$ provide us the probability of finding the excitation in the subsystem $A$ and $B$, respectively. The entanglement entropy of subsystem $A$ is obtained as $\mathcal S_A=-\rm{Tr}(\rho_A\log_2\rho_A)$ where 
 \begin{equation}
 \rho_A(t)=P_A(t)|E(t)\rangle\langle E(t)|+P_B(t)|0\rangle\langle 0|
 \label{rha}
 \end{equation}
 is the reduced density matrix of the subsystem $A$ obtained by the partial trace of $\rho_{AB}$. The state $|E(t)\rangle=\left[1/\sqrt{P_A(t)}\right]\sum_{i}^{N/2}c_i(t)|i\rangle_A$ is a general coherent single excitation state of the subsystem $A$. In Eq. (\ref{rha}) we removed the subscript $A$ from the state vectors for the convenience. Note that, the form of $\rho_A$ implies that in general, the subsystem $A$ is in a statistical mixture of  two pure states: a single excitation state ($|E(t)\rangle$) and a state with no excitations ($|0\rangle$). Diagonalizing $\rho_A$, we get two non-zero eigenvalues ($\lambda_1$ and $\lambda_2$) in complying with the two-level description, which corresponds to the probability of finding the excitation in two subsystems $A$ and $B$ [see Figs. \ref{fig:4}(b) and  \ref{fig:4}(c)]. The latter also indicates that the entropy $\mathcal S_A$ will be bounded by a maximum value of one and is independent of $N$ or $L$.  $\mathcal S_A=1$ corresponds to a maximally mixed state with equal probabilities to find the excitation either in $A$ or $B$. We assume that the excitation is initially localized in the subsystem $B$, but at the edge with subsystem $A$ i.e., $|\psi(t=0)\rangle=|0\rangle_A|N/2+1\rangle_B$. Thus, we have $\rho_A=|0\rangle\langle 0|$ and consequently $\mathcal S_A(t=0)=0$. As time progresses, the entropy increases, since the excitation diffuses into the subsystem A from B [see Fig. \ref{fig:4}(b)].  After a sufficiently long time, the system evolves into a state such that there are almost equal probabilities to find the excitation in each subsystem [see Fig. \ref{fig:4}(d)], resulting in $\mathcal S_A(t)$ approaching the value very close to one. Since the excitation was initially localized in $B$, we always have $P_B>P_A$, and the initial position of the excitation in the block $B$ determines how close $P_A$ and $P_B$ get. As expected, the larger the value of $L$, the faster the diffusion of the excitation, and it increases the growth rate of $\mathcal S_A(t)$.

Now, we consider the initial state, $|\psi(t=0)\rangle=|0\rangle_A|N/2+l_0\rangle_B$, i.e., the excitation is initially localized in the subsystem $B$, but slightly far away from the partition boundary. We see that it requires a finite time to build up the correlation between $A$ and $B$ and corresponds to the time taken by the excitation to cross the partition boundary. Fixing $l_0$, the time required to generate the correlation between $A$ and $B$ decreases with an increase in $L$, as shown in Fig. \ref{fig:4}(d) and (e). Also, for larger $l_0$, the probability of finding the excitation in $A$ at longer times gets smaller and, consequently, the correlations. A physical interpretation of the entanglement growth between $A$ and $B$ can be made based on quasiparticles \cite{cal05, sch13, jur14}. The initially localized excitation in the block $B$ is a source of quasiparticles with dispersion $\epsilon_k$, as given in Eq. (\ref{spe}). A pair of entangled quasiparticles with quasi momenta $\pm \hbar k$ propagates on either side of the lattice from the initial position of the excitation eventually makes the subsystems $A$ and $B$ to correlate.
 
%%%%%
\begin{figure*}
\centering
\includegraphics[width= 2.\columnwidth]{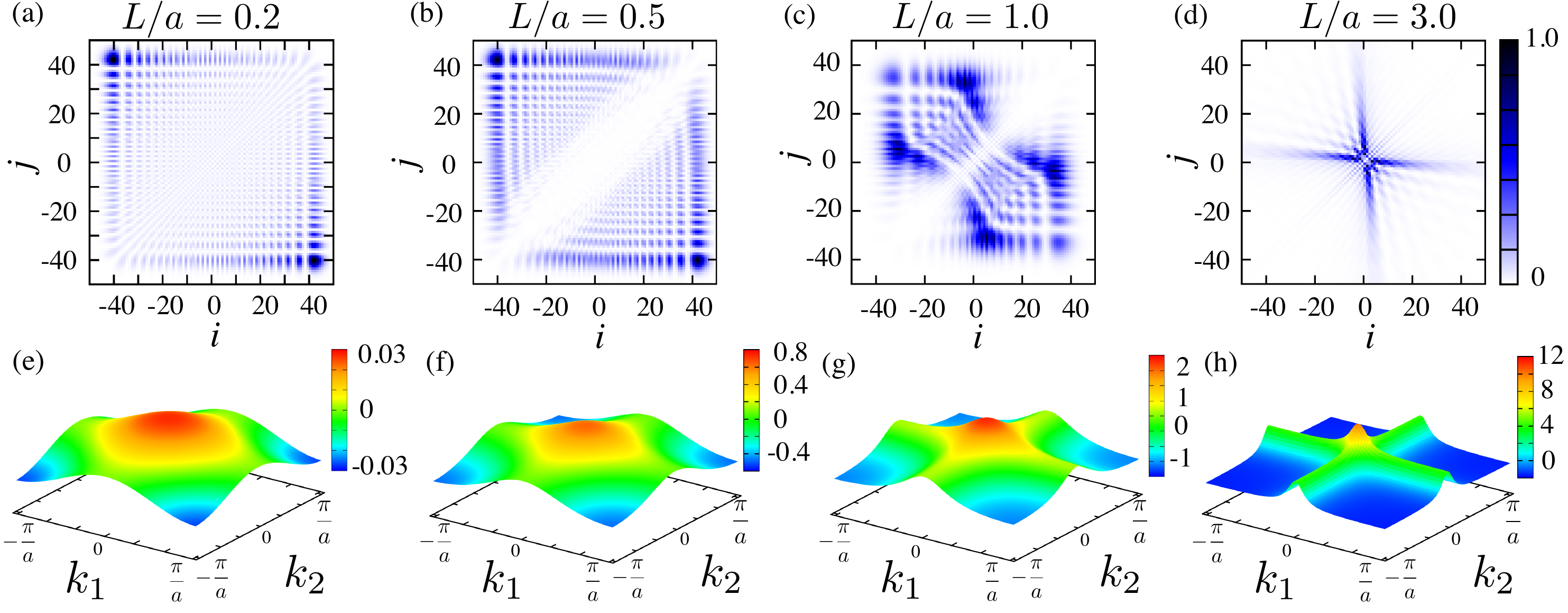}
\caption{\small{(a)-(d) The scaled two excitation distribution $\Gamma_{ij}$ for $C_6=0$ and different values of $L$. Each column is for each value of $L$. For (a) $L/a=0.2$ and (b) $L/a=0.5$, the LTPD is characterized by anti-bunching of excitations and for $L/a=3$ we have quasi-localization and tailing of excitations. For an intermediate value ($L/a=1$), the dynamics is such that one of the two excitations experiences a quasi-localization, whereas the other one propagates towards the edge of the array. The energy spectrum $\epsilon_{K,k}^0$ in $k_1-k_2$ plane for the corresponding $L$ values of (a)-(d) are shown in (e)-(h), respectively. The quasi-localization is attributed to the flat modes appear in (h) near the edges of the Brillouin zone. The times at which each of the snapshots are taken (a) $Jt=3200$, (b) $Jt=150$, (c) $Jt=30$, and (d) $Jt=15$.}}
\label{fig:5} 
\end{figure*}

%%%%%%%%%%%%%%%%%%% 
\section{Two Excitations}
\label{2ed}
The excitation dynamics becomes richer and more complex in the presence of two excitations. The Rydberg excitations interact via the van der Waals (vdW) potential $V_{ij}=C_6/r_{ij}^6$, where $r_{ij}$ is the separation between the two excitations, and $C_6$ is the vdW coefficient. First, we analyze the dynamics of non-interacting excitations ($C_6=0$), and then later extend to the interacting case ($C_6\neq 0$). Together with the RRIs, the total Hamiltonian becomes 
\begin{equation}
\hat H_t=\hat H_{ex}+\sum_{i<j}V_{ij}\hat\sigma_{ee}^i\hat\sigma_{ee}^j,
\end{equation}
with $\hat H_{ex}$ given in Eq.~(\ref{h1}). The Hamiltonian $\hat H_t$ preserves the number of excitations, and therefore we can truncate the Hilbert space into the subspace of two excitation states, i.e. $\{|ij\rangle\equiv | ... g, e^i, g, ... g, e^j, g, ...\rangle\}$, spanned by $N(N-1)/2$ states. At any instant, we have $|\psi(t)\rangle=\sum_{i<j}c_{ij}(t)|ij\rangle$ with $\sum_{i<j}|c_{ij}(t)|^2=1$ and the time-dependent probability amplitudes $c_{ij}(t)$ are obtained by solving the corresponding Schr\"odinger equations. We use scaled two-body distribution $\Gamma_{ij}(t)=|c_{ij}(t)|^2/\rm{Max}(|c_{ij}(t)|^2)$ to characterize the dynamics where Max(...) is the maximum value of $|c_{ij}(t)|^2$ at the instant $t$.

As for the case for the single excitation, the energy spectrum of $\hat H_t$ plays an important role in determining the dynamics of the two initially localized excitations \cite{qin14, cha16, let18}. To obtain the two-excitation spectrum, we introduce the center of mass $R=(i+j)a/2$ and the relative  $r=(j-i)a$ coordinates. Using $c_{ij}=\exp(iK R)\phi_K(r)$ in the Schr\"odinger equation with $\phi_K(r)= \phi_K(-r)$ and $\phi_K(0)=0$ (hard-core constraint), we obtain the eigenspectrum by solving the set of coupled equations,
\begin{equation}
\sum_{d>1}^{N-1}J_d^K \phi_K(r+ da)+ \sum_{d\neq r/a}J_d^K \phi_K(r- da)+\frac{C_6}{r^6}\phi_K(r) =\epsilon_{K,k}\phi_K(r) 
\label{req}
\end{equation}
where $J_d^K = 2Je^{-da/L}\cos(Kda/2)$ and $d$ an integer. Let $k_1$ and $k_2$ be the quasi-momenta associated with the first and second excitations, and then we have $K=k_1+k_2$ and $k=(k_1-k_2)/2$. Due to the exponential term in the expression of $J_d^K$, the parameter $L$  determines not only the range of the exchange potential but also the strength of the exchange couplings. In other words, the larger the value of $L$,  the higher the hopping matrix elements. When $C_6=0$, the states $\phi_K(r)$ are solely scattering states and the energy eigenvalues are the sum of the single-particle ones ($\epsilon_{K,k}^0=\epsilon_{k_1}+\epsilon_{k_2}$), 
\begin{equation}
\epsilon_{K,k}^0 = -J\frac{2AB-2(A+B)\cos(Ka/2)\cos(ka)+\cos Ka+\cos 2ka}{B^2 -2B\cos(Ka/2)\cos(ka)+\frac{1}{2}(\cos(Ka)+\cos(2ka))}
\label{fps}
\end{equation}
where $A=e^{-a/L}$, $B=\cosh(a/L)$ and $k$ is the relative momentum. A non-zero $C_6$ can significantly modify the features of the spectrum, in particular, the bound states emerge. The eigenspectrum is obtained by diagonalizing the Hamiltonian:
\[H^{(2)}_K=
\begin{bmatrix}
V_1+J^K_2   &  J^K_1+J^K_3  & J^K_2+J^K_4 &\dots \\
J^K_1+J^K_3 &  V_2+J^K_4    & J^K_1+J^K_5 &\dots \\
J^K_2+J^K_4 &  J^K_1+J^K_5  & V_3+J^K_6   &\dots \\
\vdots    &  \vdots       & \vdots      &\ddots\\ 
\end{bmatrix},
\]
with $V_d=C_6/(da)^6$ and $d=1, 2, 3, ...$. 
%%%%%%

\subsection{Non-interacting case ($C_6=0$)}
\label{vz}

\subsubsection{Quasi-particle spectrum and excitation dynamics}
First, we look at the non-interacting case ($C_6=0$) and analyze the dynamics as a function of both $L$ and the initial separation $d_0$ between the two excitations. Figs. \ref{fig:5}(a)-\ref{fig:5}(d) show the scaled LTPD ($\Gamma_{ij}$) for the initial condition $c_{i=N/2,j=N/2+1}=1$ in which the excitations are localized initially in adjacent sites ($d_0=1$) at the center of the array.  For  $L/a\ll 1$ [Figs. \ref{fig:5}(a) and \ref{fig:5}(b)], the exchange couplings are effectively short-range in nature, which hinders the excitations from crossing each other and resulting in the spatial anti-bunching. It is essential to point out that the anti-bunching requires a small $d_0$ because when $d_0$ gets larger the interference effects can obliterate it. We note that this anti-bunching effect also appears in the dynamics of either two hardcore bosons or two non-interacting fermions with the nearest neighbor hopping in an optical lattice \cite{qin14, beg18, lah12, fuk13, pre15,  ben12, cha16}. 

%%%%%
\begin{figure}
\centering
\includegraphics[width= 1.\columnwidth]{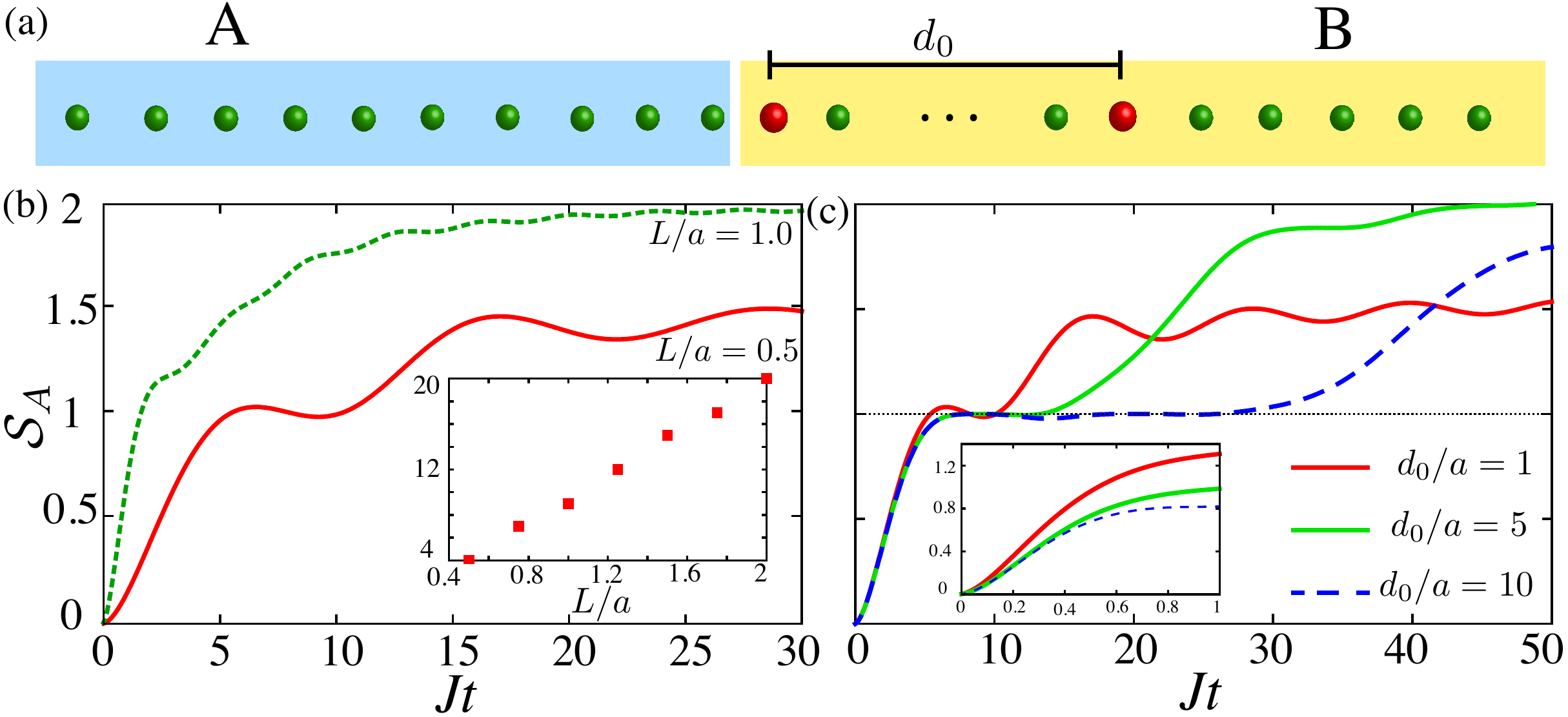}
\caption{\small{(a) The schematic of the initial state in which the two excitations (red spheres) are localized in the subsystem $B$ at a separation of $d_0$. (b) The dynamics of $\mathcal S_A$ for $d_0=a$ with $L/a=0.5$ (solid line) and $L/a=1$ (dashed line). The inset shows the number of non-zero eigenvalues of $\rho_A$ as a function of $L$ at $Jt=15$ and it increases linearly with $L$. (c) The dynamics of $\mathcal S_A$ for $L/a=0.5$ with different $d_0$. The thin horizontal line stands for $\mathcal S_A=1$. The inset shows the initial growth of $S_A(t)$ for different $d_0$.}}
\label{fig:6} 
\end{figure}

As $L$ increases, $\epsilon_{K,k}^0$ gets modified as shown in Figs. \ref{fig:5}(e)-\ref{fig:5}(h), where $\epsilon_{K,k}^0$ is plotted in the $k_1-k_2$ plane, and this affects the excitation dynamics. For large values of $L$, we observe quasi-localization of excitations [see Fig. \ref{fig:5}(d)]. This is identical to that of the single excitation discussed in Sec. \ref{sed} and both excitations are more favored to be found at their initial positions. The quasi-localization at large $L$ is attributed to those modes having both $\partial \epsilon_{K,k}^0/\partial {k_1}\sim 0$, and $\partial \epsilon_{K,k}^0/\partial {k_2}\sim 0$. They appear as flat modes near $k_1, k_2\sim\pm\pi/a$, and the local maxima of the emerging energy stripes, as shown in Fig. \ref{fig:5}(h). For the intermediate value of $L$, we have a scenario in which one of the two excitations experiences a quasi-localization, and the other one propagates towards the edge of the array, as seen in Fig. \ref{fig:5}(c). The latter arises from those modes in the stripes with either $\{\partial \epsilon_{K,k}^0/\partial {k_1}\sim 0, \partial \epsilon_{K,k}^0/\partial {k_2}\neq 0\}$ or vice versa. The same holds for the tailing behavior in Fig.~\ref{fig:5}(d), which shows almost horizontal and vertical probability tails in the LTPD. As the initial separation between the two excitations ($d_0$) increases, the LTPD reveals us non-trivial patterns due to the quantum interference, especially at small values of $L$. The corresponding results for the LTPD ($\Gamma_{ij}$) are shown in Appendix \ref{a2}. 
%%%%%
\subsubsection{Entanglement entropy}
\label{ee2}
The presence of the second excitation enhances the bipartite entanglement between sublattices $A$ and $B$. The Hilbert space of each subsystem is spanned by states with zero, one and two excitations, i.e., $\mathcal H_A\in \{|0\rangle_A, |i\rangle_A, |ij\rangle_A\}$ and $\mathcal H_B\in \{|0\rangle_B, |i\rangle_B, |ij\rangle_B\}$, where $i$ and $j$ represent the indices of the site in which the excitations reside in each subsystem. In this basis, the general state of the system can be rewritten as
\begin{equation}
\ket{\psi} = \sum_{i<j}^{N/2} c_{ij} \ket{ij}_A \ket{0}_B+  \sum_{i=1}^{N/2} \sum_{j=N/2+1}^{N} c_{ij}\ket{i}_A\ket{j}_B+\sum_{N/2<i<j}^{N} c_{ij} \ket{0}_A \ket{ij}_B.
\end{equation}
We have the following three scenarios: two excitations in $A$, two excitations in $B$, or one excitation in each of the subsystems. We look at the growth of the entanglement entropy of the subsystem $A$ by assuming both the excitations are initially placed in $B$, and in particular, one is localized in the adjacent site to the partition boundary [see Fig. \ref{fig:6}(a)]. 
As expected, for larger $L$, the growth of $\mathcal S_A(t)$ becomes faster as shown in Fig. \ref{fig:6}(b) for $d_0=1$. At longer times, we observe that $\mathcal S_A(t)$ saturates and oscillates around a steady value. Unlike that for the single excitation case discussed in Sec. \ref{seee}, for two excitations, the maximum or the long time value of $\mathcal S_A(t)$ at longer times, is not globally bounded but depends on the system parameters $L$ and $d_0$. This behavior is understood as follows: increasing the value of $L$ enlarges both the range and strength of exchange couplings, which in turn makes the states $|ij\rangle$ with different $d_0$ to be energetically separated. The latter results in the appearance of more non-zero eigenvalues for $\rho_A$ at longer times [see inset of Fig. \ref{fig:6}(b)], and consequently higher values for $\mathcal S_A$ at larger $L$.

\begin{figure}
\centering
\includegraphics[width= .9\columnwidth]{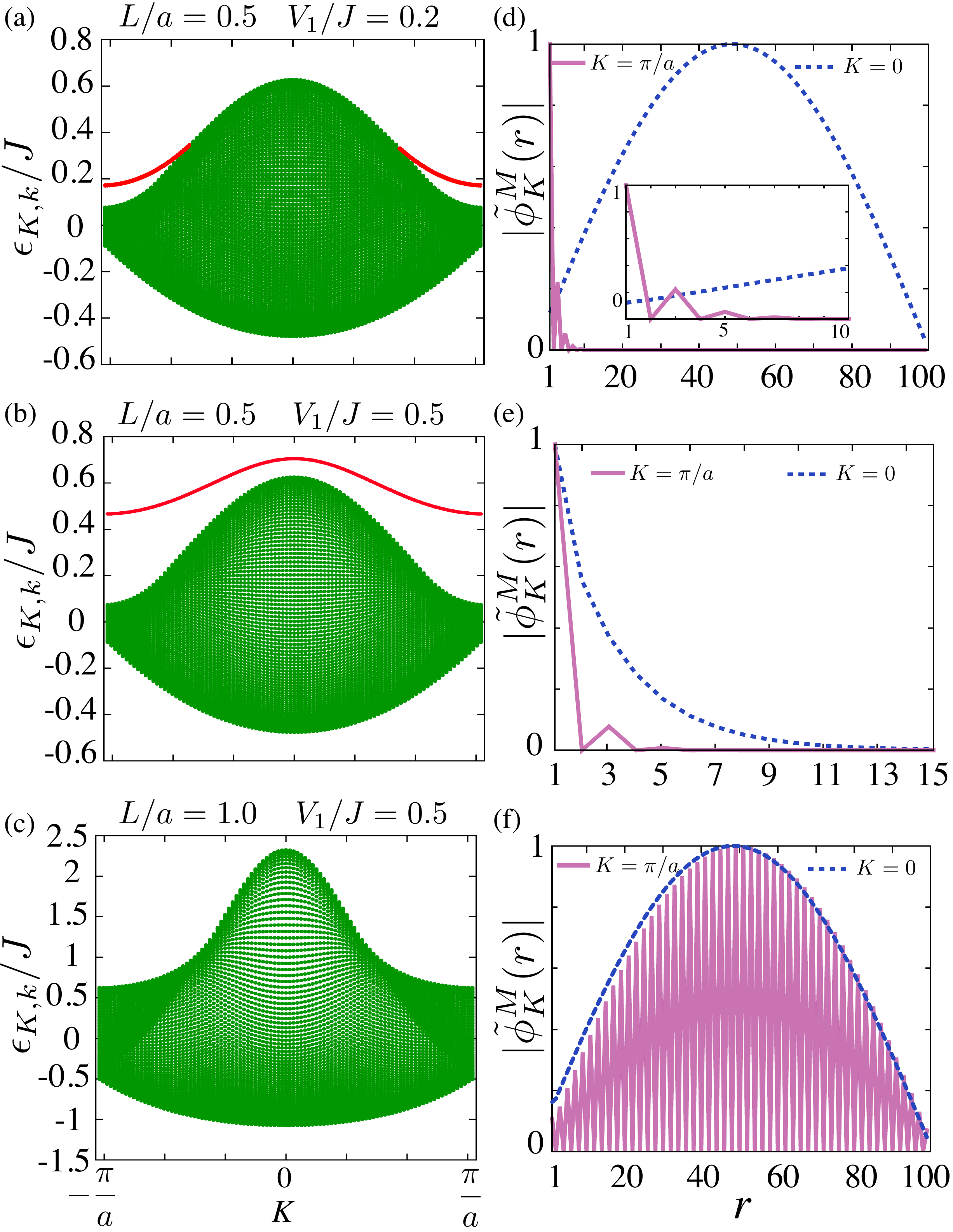}
\caption{\small{(Color online) The two-particle excitation spectrum for (a) $L/a=0.5$, $V_1/J=0.2$, (b) $L/a=0.5$, $V_1/J=0.5$ and (c) $L/a=1$, $V_1/J=0.5$. The solid (red) line in (b) shows the isolated band for the bound states. (d)-(f) show the highest radial eigenfunctions (scaled by its maximum value) for $K=0$ (dashed line) and $K=\pi/a$ (solid line), respectively for the excitation spectrum shown in (a)-(c).}}
\label{fig:7} 
\end{figure}

In Fig. \ref{fig:6}(c), we show the dynamics of $\mathcal S_A$ for $L/a=0.5$ and various $d_0$, where the anti-bunching dynamics occurs between the excitations for $d_0=1$. The initial growth of $\mathcal S_A$ in time is independent of $d_0$  [see the inset of Fig. \ref{fig:6}(c)], and it is mainly originated by the excitation closer to the partition boundary. For $d_0=1$, it is the anti-bunching, which partially nullifies the contribution of the second excitation to $\mathcal S_A$ at the initial stage of the dynamics, whereas, for larger $d_0$, it is the initial separation itself. Eventually, $\mathcal S_A$ exceeds $1$, signaling the role of the second excitation.  The second excitation takes a finite time to diffuse into the subsystem $B$, which depends on $d_0$ and $L$. This results in the freezing of $\mathcal S_A$ around $\mathcal S_A\sim 1$ for a limited time. For a fixed $L$, the freezing time increases with $d_0$, as seen in Fig. \ref{fig:6}(c) and also decreases with $L$ for a fixed $d_0$. All of these imply that it is possible to coherently control the bipartite entanglement, both its dynamics and long time steady values,  by simply varying the initial separation between two excitations.

\subsection{Rydberg interacting case ($C_6\neq 0$)}
\subsubsection{Bound and Scattering States}

%%%%%%%%%%
\begin{figure}
\centering
\includegraphics[width= .9\columnwidth]{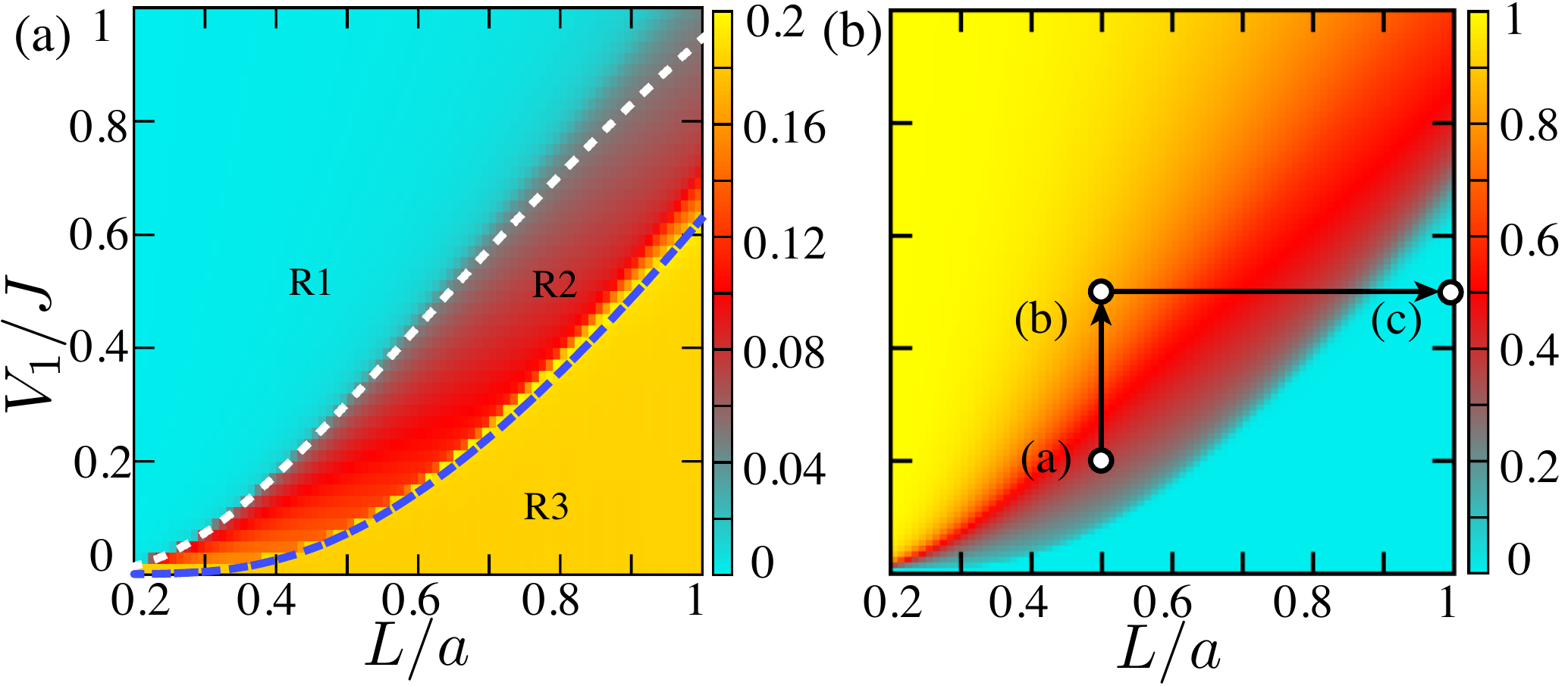}
\caption{\small{(a) The scaled, $K$-averaged $\Delta r$ as a function of $V_1$ and $L$. In region R1, the excitation spectrum $\epsilon_{K,k}$ is characterized by an isolated band of bound states, in region R2 spectrum consists of at least one bound state, but there is no isolated band for bound states [see Fig. \ref{fig:7}(a)], and in R3 region, $\epsilon_{K,k}$ has only scattering states. The dashed line between regions R1 and R2 is obtained from a truncated model whereas the same between regions R2 and R3 is provided by the criteria $V_1^{cr}=\epsilon_{\pi/a, \pi/2a}^0$. (b) The $K$-integrated overlap $\mathcal O_1^M$ of $\phi_{K}^M(r)$ on the initial state in which the two excitations are localized at the nearest neighbor sites ($d_0=1$). The values of $V_1$ and $L$ at a, b, and c for which the two-excitation dynamics is shown in Figs. \ref{fig:10}(a)-\ref{fig:10}(c).}}
\label{fig:8} 
\end{figure}

For sufficiently large values of $C_6$ or $V_1=C_6/a^6$, a new energy band starts to emerge and eventually separates from the scattering states at a higher $V_1$, as shown in Figs. \ref{fig:7}(a) and \ref{fig:7}(b). The new isolated band in Fig. \ref{fig:7}(b) corresponds to the interaction induced bound states of two excitations. These bound states are identical to the magnon bound states in spin models \cite{fuk13, let18} or the particle-bound states in Hubbard models \cite{win06, pii07, pet07, val08, val09}. They are characterized by exponentially decaying $\phi_K(r)$ with the maximum at $r=1$, showing that it is morelikely to find the excitations at adjacent sites. In contrast, the scattering states are delocalized over all lattice sites. Thus, the quasi-momentum $K$ is associated with two kinds of quasi-particles depending on the nature of the eigenstates of $\phi_K(r)$, scattering or bound state quasi-particles. As we will show below, RRIs favor the existence of bound states, whereas the exchange couplings suppress them. To demonstrate the competition between RRIs and exchange couplings, in Figs. \ref{fig:7}(d)-\ref{fig:7}(f), we show the topmost (corresponding to the largest $\epsilon_{K, k}$) radial eigenfunctions $\phi_K^M(r)$, in the case of $K=0$ (dashed line) and $K=\pi/a$ (solid line) for different values of $L/a$ and $V_1/J$ \footnote{We have scaled the RRI strength by $J$ to leave the $L$-dependence only in the exponential function in the Hamiltonian $\hat H_{ex}$. Notice that to compare this interaction with the RRI strength $V_1$ it is more adequate to use the re-scaled nearest neighbor hopping strength $J'=J\exp(-a/L)$", which also requires to re-scale $f'=f\exp(a/L)$.}. For $L/a=0.5$ and $V_1/J=0.2$, the function $\phi_{K=\pi/a}^M(r)$ is a bound state, but $\phi_{K=0}^M(r)$ is a scattering state, as shown in Fig. \ref{fig:7}(d). Keeping  $L/a=0.5$ fixed and by increasing $V_1/J$ from 0.2 to 0.5, the radial eigenfunction at $K=0$ also becomes a bound state, resulting in an isolated band of bound states [see Figs. \ref{fig:7}(b) and \ref{fig:7}(e)]. Besides, we found that the bound state at $K=\pi/a$ is more localized than the one at $K=0$ for a given $L$ and $V_1$. Now keeping $V_1/J=0.5$ and increasing $L/a$ to 1, the exchange couplings dominate RRIs, which eliminates the bound states from the spectrum [see Figs. \ref{fig:7}(c) and \ref{fig:7}(f)].

To characterize the competition between the RRIs and exchange couplings it is convenient to introduce a new parameter, the scaled radial width of the functions $\phi_K^M(r)$, and averaged over $K$, 
\begin{equation}
\Delta r=\frac{a}{2\pi N}\int_{-\pi/a}^{\pi/a}\Delta r_K dK,
\end{equation}
where $\Delta r_K=\sqrt{\langle r^2\rangle_K-\langle r\rangle_K^2}$, and the symbol $\langle ...\rangle_K$ indicates that the average is taken over the state $\phi_K^M(r)$. In Fig. \ref{fig:8}(a), we show $\Delta r$ as a function of $V_1$ and $L$, which identifies three different regions R1, R2, and R3. For sufficiently small values of $L/a$ and large values of $V_1/J$ (region R1), we have $\Delta r\sim 0$. The latter implies that the radial states $\phi_K^M(r)$ are purely bound states, highly localized, and the excitation spectrum has an isolated band of bound states, as shown in Fig. \ref{fig:7}(b). At the other end, in region R3, the width $\Delta r$ takes the maximum value ($\sim 0.2$). The latter means that $\phi_K^M(r)$ are purely scattering states or an excitation spectrum with no bound states, as shown in Fig. \ref{fig:7}(c). The intermediate region R2 stands for an excitation spectrum similar to the one shown in Fig. \ref{fig:7}(a), i.e., there are bound states but not yet developed as a completely isolated band.

Now we focus on obtaining analytical expressions for the boundaries separating the regions R1, R2, and R3. To simplify the analysis, we truncate the exchange couplings up to next nearest neighbor and the RRIs up to the nearest neighbor terms. We obtain the energy of scattering and bound states as \cite{let18}, 
\begin{eqnarray}
&&\epsilon_{K, k}^0\simeq 2\left(J_1^K\cos(ka)+J_2^K\cos(2ka)\right), \\
&&\epsilon_{K}^b\simeq 2J_2^K+\frac{\left(J_1^K\right)^2}{V_1}+\frac{\left(J_1^K\right )^2J_2^K}{V_1^2}+\frac{V_1^2}{V_1+J_2^K},
\end{eqnarray}
respectively. The term $J_d^K$ is given below Eq. (\ref{req}) and the bound-state energy is independent of $k$. A bound state exists if 
\begin{equation}
\epsilon_{K}^b-(\epsilon_{K, k}^0)_{max}>0,
\label{cri}
\end{equation}
 where $(\epsilon_{K, k}^0)_{max}$ is the maximum eigenvalue among the scattering states for a given $K$. Since the bound state first appears at $K=\pi/a$ the boundary between the regions R2 and R3 in Fig. \ref{fig:8}(a) is given by $\epsilon_{\pi/a}^b-(\epsilon_{\pi/a, k}^0)_{max}=0$. The latter provides us the critical RRI strength $V_1^{cr}$ above which we have at least a single bound state in the energy spectrum. Analytically, we have $V_{cr}^1=4Je^{-2a/L}$ and also using the truncated model, we have $\epsilon_{\pi/a, \pi/2a}^0\simeq 4Je^{-2a/L}$. Therefore we assume that the critical value, $V_{cr}^1=\epsilon_{\pi/a, \pi/2a}^0$ is valid beyond the truncated model and this is found to be in good agreement with the numerical results even for sufficiently large $L$. Similarly, by taking $K=0$ in Eq. (\ref{cri}), we can estimate a second critical RRI strength $V_{cr}^2$ above which we have a bound state at $K=0$ or equivalently to have an isolated band of bound states. The truncated model gives us $V_{cr}^2\simeq \left [J_1^0+\sqrt{(J_1^0)^2+4J_1^0J_2^0}\right]/2$ with $J_1^0=2Je^{-a/L}$ and $J_2^{0}=2Je^{-2a/L}$, and this value separates the regions R1 and R2 in Fig. \ref{fig:8}(a),  At large values of $L$, the analytical estimation of $V_{cr}^2$ from the truncated model starts to deviate from the exact numerical results since the long-range nature of the interactions become very significant. Henceforth, we restrict to $L/a\in [0,1]$ with no restrictions on $V_1$.

%%%%%%%%%%
\begin{figure}[hbt]
\centering
\includegraphics[width= 1.\columnwidth]{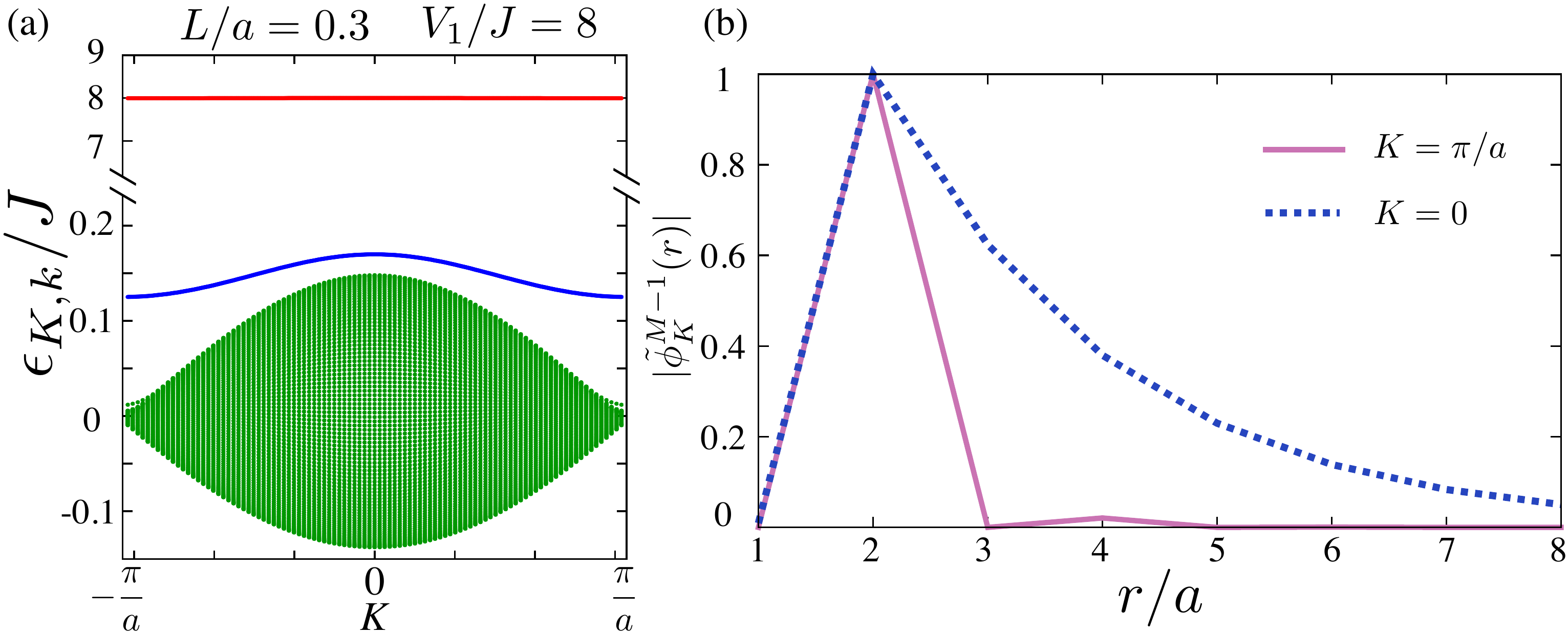}
\caption{\small{(a) The quasi-particle spectrum with two bands of bound states. The top most one we term it as the zeroth order bound states, with radial functions $\{\phi_K^{M}(r)\}$ having a peak at $r=a$. The second highest band of bound states are represented by $\{\phi_K^{M-1}(r)\}$ having a peak at $r=2a$ as shown in (b) $K=0$ (dashed line) and $K=\pi/a$ (solid line).}}
\label{fig:9} 
\end{figure}

{\em Strongly interacting case}.--- For $V_1\gg V_{cr}^2$, bands of higher-order bound states emerge in the spectrum. We define an $n$th order bound states as those having the radial wavefunctions $\{\phi_K^{M-n}(r)\}$ with a peak at $r=(n+1)a$. We call the band of bound states appearing in Fig. \ref{fig:7}(b) as the zeroth-order one. A spectrum exhibiting both zeroth and first order bound states is shown in Fig. \ref{fig:9}(a). The radial eigenfunctions $\phi_{\pi/a}^{M-1}(r)$ and $\phi_{0}^{M-1}(r)$ are shown in Fig. \ref{fig:9}(b), and they both exhibit a peak at $r=2a$. As we will discuss below, the presence of higher-order bound states in the eigenspectrum lead to non-trivial dynamical scenarios for a given initial state.

%%%%%%%%%%%%%
\subsubsection{Dynamics}
We find that the dynamics critically depends on the initial state overlap with the Hamiltonian eigenstates. We define an overlap function, 
\begin{equation}
\mathcal O_{d_0}^M=\int_{-\pi/a}^{\pi/a}|\langle \phi_K^M|\psi(t=0)\rangle|^2 dK,
\label{of}
\end{equation}
which measures the amount of bound state quasi-particles from the zeroth-order band for an initial state of two localized excitations separated by a distance of $d_0 a$. In Fig. \ref{fig:8}(b), we show $\mathcal O_{d_0}^M$ for $d_0=1$ as a function of $V_1$ and $L$. For $V_1<\epsilon_{\pi/a, \pi/2a}^0$ or at the region R3 in Fig. \ref{fig:8}(a), no bound states exist, and consequently $\mathcal O_1^M$ is vanishingly small. In other words, the initial state is only a source of scattering state quasi-particles. At the other extreme, i.e., for $V_1\gg V_{cr}^2$, we have $\mathcal O_1^M\sim 1$, indicating that the initial state is only a source of bound state quasi-particles. In the intermediate regime, the initial state is a source of both scattering and bound-state quasi-particles. As we see below, the dynamics crucially depends on the how much fraction of bound-states exists in the initial state.

%%%%%%%%%%
\begin{figure}[hbt]
\centering
\includegraphics[width= .9\columnwidth]{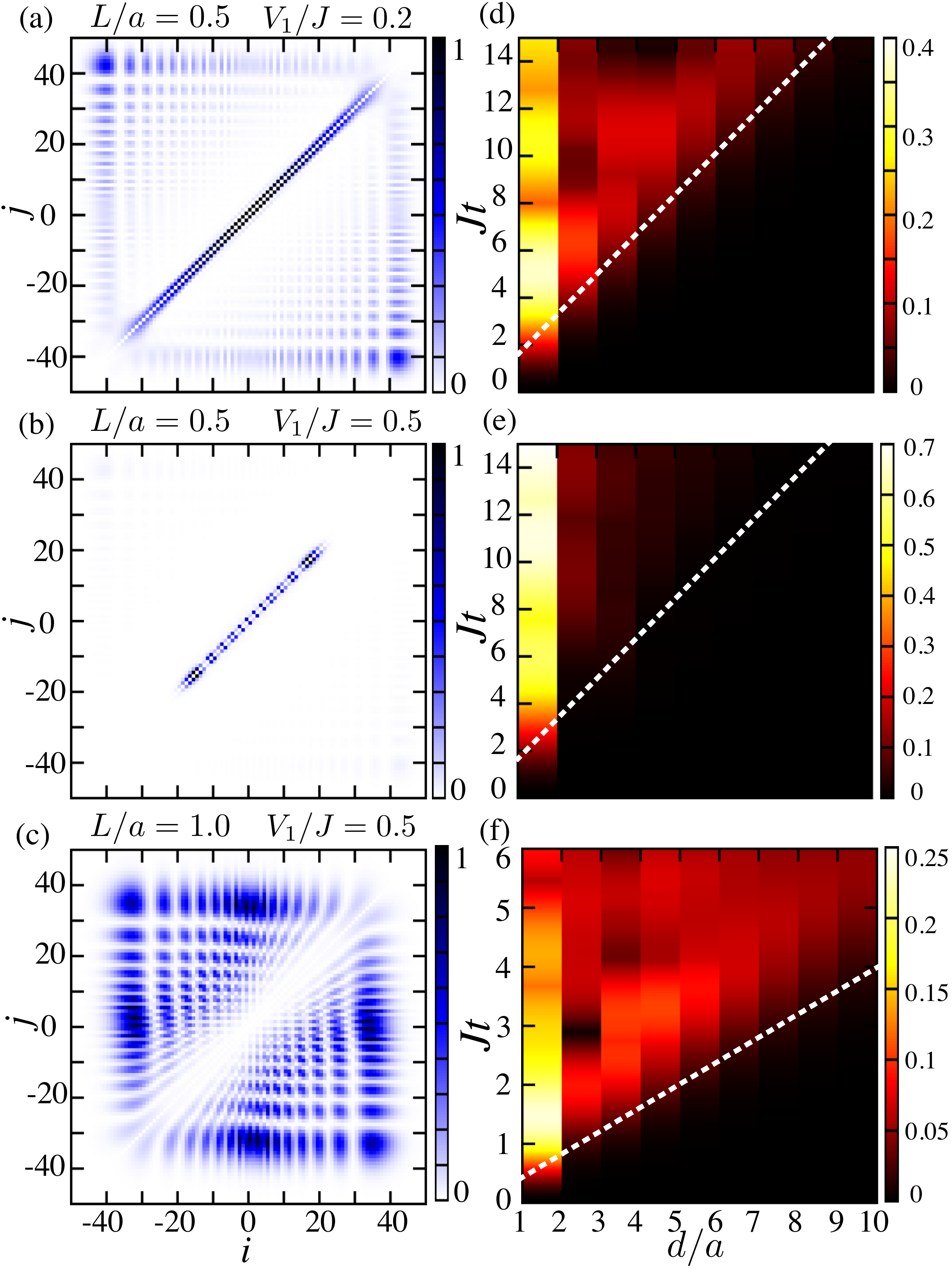}
\caption{\small{(a)-(c) show the two-excitation dynamics [$\Gamma _{ij}(t)$] for the initial state in which the two excitations are localized in the nearest neighbor sites ($d_0=1$) at the center of the lattice. The interaction parameters are given above each plot, and the excitation spectrum for the same parameters are shown in Fig. \ref{fig:7}(a)-\ref{fig:7}(c). The plots (d)-(f) show the corresponding correlations $C(d,t)$ for the dynamics shown in (a)-(c), respectively. The dashed line in (d)-(f) shows the theoretical estimate of maximum group velocity [Eq. (\ref{mv})] from the single excitation spectrum. The times at which each of the snapshots are taken (a) $Jt=150$, (b) $Jt=150$, and (c) $Jt=30$.}}
\label{fig:10} 
\end{figure}

In Figs. \ref{fig:10}(a)-\ref{fig:10}(c), we show the dynamics of two excitations initially localized at the nearest neighbor sites  ($d_0=1$), for the same parameters as in Figs. \ref{fig:7}(a)-\ref{fig:7}(c), respectively.  For $L/a=0.5$ and $V_1/J=0.2$ [a point in region R2 in Fig. \ref{fig:8}(a), and also in Fig. \ref{fig:8}(b) marked as (a)] the initial state has contributions from both  bound and scattering states. The latter results in two features in the dynamics, as seen in Fig. \ref{fig:10}(a): the diagonal stripes are indicating the quantum walk of a bound pair of excitations, and the anti-bunching of excitations due to the scattering states. Keeping $L/a=0.5$, and increasing $V_1/J$ to $0.5$  [marked as (b) in Fig. \ref{fig:8}(b)], the dynamics becomes a quantum walk of the bound state of two excitations predominantly, as shown in Fig. \ref{fig:10}(b). This behavior is expected from the nature of the excitation spectrum, which exhibits an isolated band of bound states [see Fig. \ref{fig:7}(b)], and the overlap parameter $\mathcal O_1^M$ is very close to 1. Now, keeping $V_1/J$ fixed to $0.5$ and increasing $L/a$ to $1.0$ [marked as (c) in Fig. \ref{fig:8}(b)], we observe in Fig.~\ref{fig:10}(c) that the bound state features completely disappear from the dynamics [see Fig. \ref{fig:10}(c)], in agreement with the absence of bound state eigenstates from the excitation spectrum shown in Fig. \ref{fig:7}(c).  In this way, the dynamics also manifests the competition between RRIs and exchange couplings.
%%%%%%%

Further, we look at the propagation of two-point correlations in the excitation dynamics, 
\begin{equation}
C(d, t)=\sum_i\left(\langle \hat \sigma_{ee}^i\hat \sigma_{ee}^{i+d}\rangle-\langle \hat\sigma_{ee}^i\rangle\langle\hat \sigma_{ee}^{i+d}\rangle\right)
\end{equation}
where $\langle ... \rangle\equiv \langle\psi(t)| ...|\psi(t)\rangle$. We display the dynamics of $C(d,t)$ in Figs. \ref{fig:10}(d)-\ref{fig:10}(f), for the same parameters used in Figs. \ref{fig:10}(a)-\ref{fig:10}(c), respectively.  It can be seen that if the initial state has sufficient overlap with the scattering states, the correlations exhibit an effective light-cone like behavior [see Figs. \ref{fig:10}(d) and \ref{fig:10}(f)] \cite{lau08, mar12, jur14,phi14}.  This means that the correlations decay exponentially beyond a causal region. Such a Lieb-Robinson upper bound is known to be a feature of short-range \cite{lie72} or weakly long-range interacting quantum systems \cite{hau13}. Interestingly, even for sufficiently long-range hopping ($L/a=1$), we have a light-cone behavior in the correlations [see Fig. \ref{fig:10}(f)]. The upper bound for the speed [see dashed lines in Figs. \ref{fig:10}(d) and \ref{fig:10}(f)] at which the correlations propagate can be estimated from the quasi-particle spectrum of the single excitation, given in Eq. (\ref{spe}). Analytically, we obtain 
\begin{equation}
v_g^{max}=\left(\frac{d\epsilon_k}{dk}\right)_{max}=-Ja \frac{\sinh(a/L)\sin (qa)}{\left[\cosh (a/L) - \cos (qa)\right]^2},
\label{mv}
\end{equation}
where $$q=\frac{1}{a}\cos ^{-1} \left[\frac{1}{2}\left(\sqrt{\cosh^2(a/L)+8}-\cosh(a/L)\right)\right].$$ On the other hand, if the initial state is dominated by bound state quasiparticles (in the limit of sufficiently large $V_1$), the propagation of correlations are significantly slowed down, and one observes a long-surviving peak at $d=a$, as seen in Fig. \ref{fig:10}(e). Effectively we have a quantum walk of a bound pair of excitations, as shown in Fig. \ref{fig:10}(b). 
%%%%%%%%%%%%%%%%%%%%%%%
%%%%%%%%%%
\begin{figure}
\centering
\includegraphics[width= 1.\columnwidth]{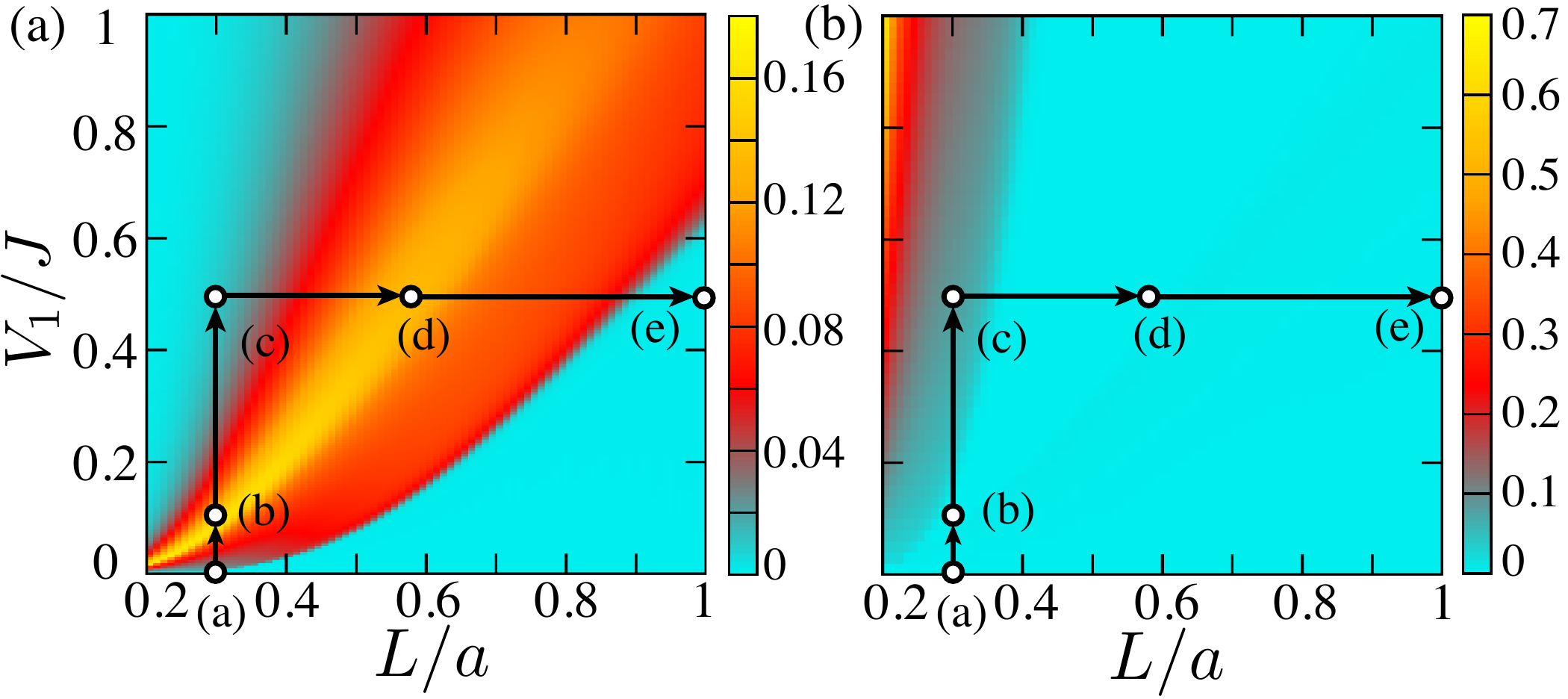}
\caption{\small{The initial state overlaps (a) $\mathcal O_{2}^{M}$  and (b) $\mathcal O_{2}^{M-1}$ for $d_0=2$. For the points marked as a, b, c, d, and e, the dynamics and correlations are shown in Fig. \ref{fig:12}.}}
\label{fig:11} 
\end{figure}

%%%%%%%%%%
\begin{figure}
\centering
\includegraphics[width= .9\columnwidth]{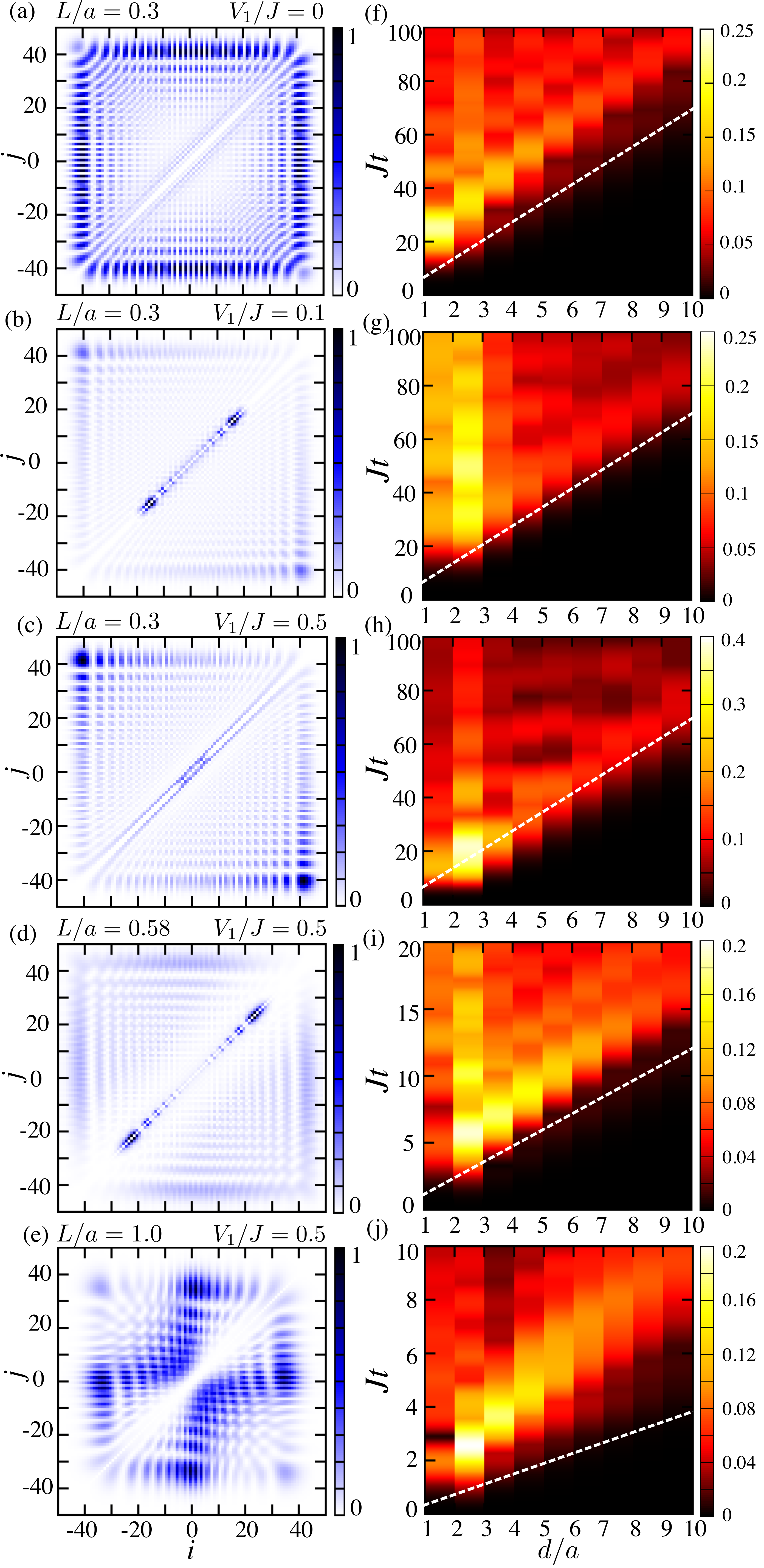}
\caption{\small{(a)-(e) show the dynamics with $d_0=2$ for the points (a)-(e) marked in Fig. \ref{fig:11}. The corresponding dynamics of the correlations $C(d,t)$ are shown in (f)-(j), respectively. The dashed line shows the theoretical estimate of maximum group velocity [Eq. (\ref{mv})] from the single excitation spectrum. In (g) and (i), the bound state character is revealed by the prominent peak along $d=2$. In all these cases, we see a light-cone like structure in the correlation dynamics. The times at which each of the snapshots are taken (a)-(c) $Jt=600$, (d) $Jt=110$, and (e) $Jt=30$.}}
\label{fig:12} 
\end{figure}

{\em Effect of the initial separation between excitations}.---The initial separation $d_0$ of the two localized excitations also has a significant impact on the dynamics. The interplay between $V_1$ and $L$ become more complex for larger initial distances between the excitations, and we restrict the discussions to the case of $d_0=2$. For $d_0=1$, we have seen that the overlap function $\mathcal O_1^M$ was able to capture the physics completely, whereas, for $d_0=2$, we need both $\mathcal O_2^M$ and $\mathcal O_2^{M-1}$, which measures the fraction of first-order and second-order bound state quasi-particles in the initial state, respectively. In contrast to both $\mathcal O_1^M$ and $\mathcal O_2^{M-1}$, the overlap quantity $\mathcal O_2^M$ exhibits a non-monotonous behavior as a function of both $L$ and $V_1$ [see Fig. \ref{fig:11}]. This non-monotonous behavior is directly linked to the non-zero values of $\mathcal O_2^{M-1}$, especially at large $V_1$, and below, we analyze how it affects the excitation dynamics.

The excitation and two-point correlation dynamics as a function of $V_1$ and $L$ for $d_0=2$ are shown in Fig.~\ref{fig:12}. For $d_0=1$, we have seen that the presence of bound states in the dynamics monotonously increases with $V_1$ while keeping $L$ constant or decreases with $L$ for a fixed $V_1$. On the contrary, for $d_0=2$ (see Fig. \ref{fig:12}), the two excitation dynamics shows a non-monotonous behavior as a function of both $L$ and $V_1$. To exemplify the latter, we show the dynamics keeping $L/a=0.3$ and varying $V_1/J$. For $V_1/J=0$, the dynamics involves no bound states [see Fig. \ref{fig:12}(a)], but when increasing the RRIs up to $V_1/J=0.1$, the dynamics exhibits prominent bound state character [see Fig. \ref{fig:12}(b)]. For $L/a=0.3$ and $V_1/J=0.1$ [see point (b) in Figs. \ref{fig:11}(a) and \ref{fig:11}(b)], we have $\mathcal O_2^M\neq 0$ and $\mathcal O_2^{M-1}\sim 0$, which implies that the bound states shown in Fig. \ref{fig:12}(b) is of the first-order type. However, if we increase $V_1/J$ further, the contribution from the bound states reduces and that from the scattering states enhances [see Fig. \ref{fig:12}(c)]. That is because $\mathcal O_2^M$ reduces, and $\mathcal O_2^{M-1}$ hardly gain by that increment in $V_1$ [marked as point (c) in Figs. \ref{fig:11}(a) and \ref{fig:11}(b)]. Note that the small fraction of bound states presents in Fig. \ref{fig:12}(c) is a superposition of first and second-order bound states. On the other hand, if we keep $V_1/J=0.5$ but increase $L/a$ to a value of 0.58, we found the counter-intuitive effect that the bound state contribution gets enhanced in the dynamics [see Fig. \ref{fig:12}(d)]. Nevertheless, a further increment in $L/a$ eliminates the bound state character, as seen in Fig. \ref{fig:12}(e) for $L/a=1$. For all these cases, we see a light-cone like structure in the propagation of two-point correlations, as shown in Figs.~\ref{fig:12}(f)-\ref{fig:12}(j), due to presence of scattering states.

\begin{figure}
\centering
\includegraphics[width= .9\columnwidth]{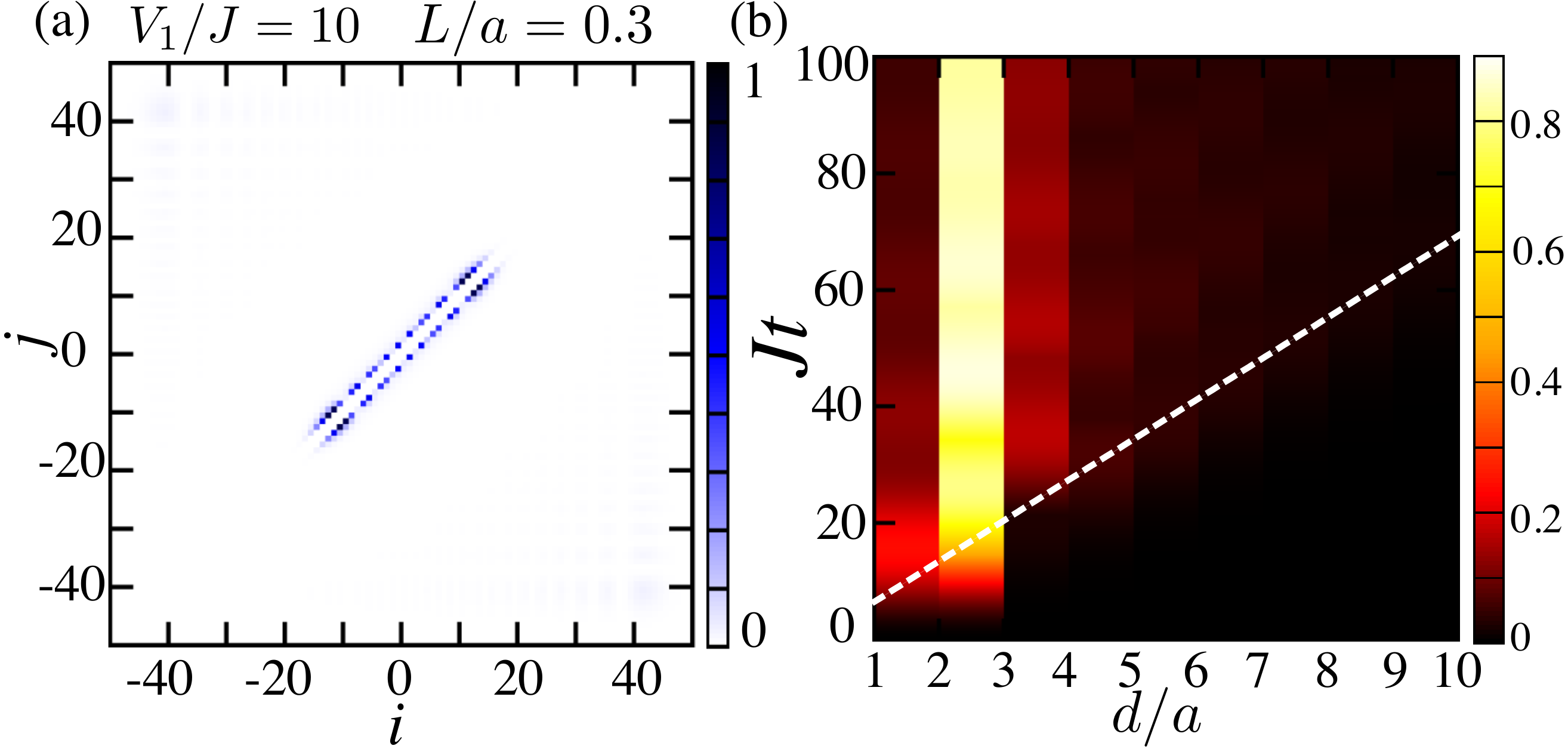}
\caption{\small{The dynamics of (a) excitations and (b) correlations $C(d, t)$ with $d_0=2$. The vertical stripe of maximum in $C(d,t)$ along $d=2a$ indicating the presence of bound state quasiparticles with radial functions  $\phi_K^{M-1}(r)$. The snapshot is taken at $Jt=600$.}}
\label{fig:13} 
\end{figure}

To conclude this discussion, we look at the dynamics for $d_0=2$ with significantly large $V_1$ such that in the excitation spectrum, we have two isolated bands of bound states. In this case, the dynamics is characterized by the quantum walk of a bound state with two excitations separated by one vacant site (see Fig. \ref{fig:13}), i.e., the second-order bound state. In that case, the propagation of the two-point correlations is significantly halted by the bound state as shown in Fig.~\ref{fig:13}(b).

%%%%%%%%%%

%%%%%%%%%%%%%%%%%%%%%%%
\subsubsection{Entanglement entropy}
\begin{figure}[hbt]
\centering
\includegraphics[width= 1.\columnwidth]{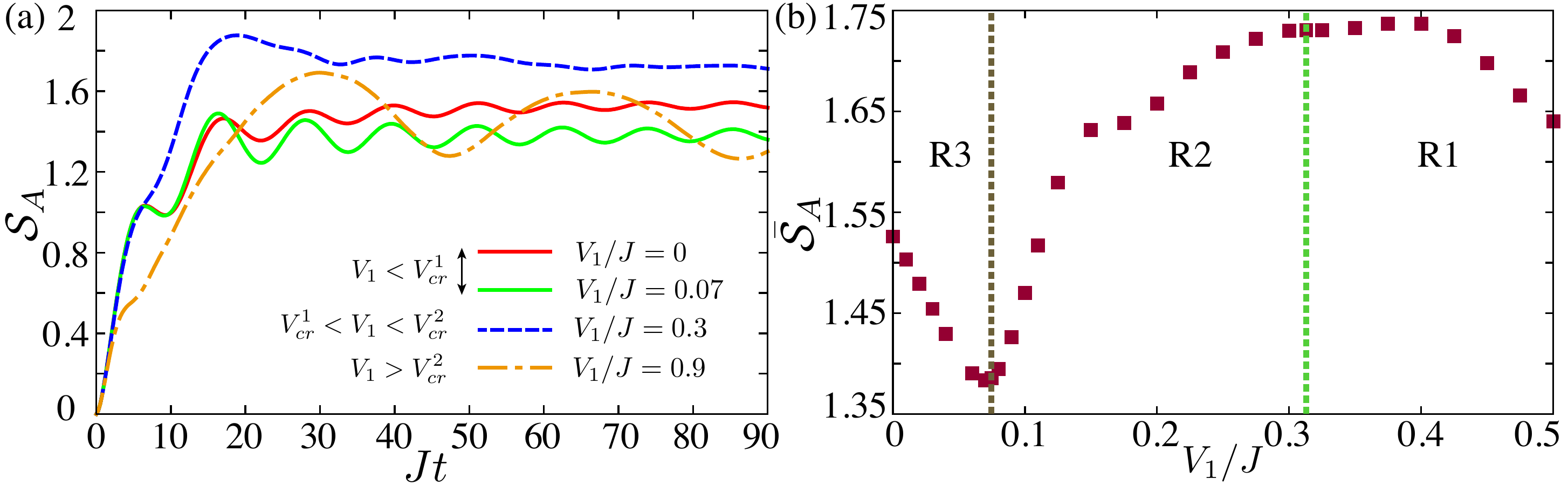}
\caption{\small{The dynamics of $\mathcal S_A(t)$ for $d_0=1$ and $L/a=0.5$ with different values of $V_1$. The values of $V_1$ covers three different regions (R1, R2, and R3) based on the nature of excitation spectrum. In R3, we have $V_1<V_{cr}^1$, and the excitation spectrum possesses no bound states, and the effect of interaction is to reduce the effective exchange couplings. In R2 ($V_{cr}^1<V_1<V_{cr}^2$), excitation spectrum has a partially separated band of bound states and, in R1 ($V_1>V_{cr}^2$) the spectrum possesses an isolated band of bound states. Deep in R1, the initial state becomes a source of bound state quasi-particles. (b) shows the behavior of $\bar {\mathcal S}_A$ in three different regions, and dashed vertical lines show $V_{cr}^1$ (left) and $V_{cr}^2$ (right).}}
\label{fig:14} 
\end{figure}

In this section, we extend the analysis of $\mathcal S_A(t)$, to include the effect of RRIs. In particular, we look at the impact of interaction induced bound states on the growth of bipartite entanglement. Again, we assume that the two excitations are initially localized in the subsystem $B$, as shown in Fig. \ref{fig:6}(a), and use the basis states having zero, one, and two excitations, as discussed in Sec.~\ref{ee2}. We see that RRIs do not affect the growth of $\mathcal S_A(t)$ qualitatively but only quantitatively. To show this, we fix $L/a=0.5$ (for which $V_{cr}^1/J\sim 0.073$ and $V_{cr}^2\sim 0.303$) and vary $V_1$, see Fig.~\ref{fig:14}(a). Again, we consider the three different regions: R3 ($V_1<V_{cr}^1$), R2 ($V_{cr}^1<V_1<V_{cr}^2$), and R1 ($V_1>V_{cr}^2$) as in Fig. \ref{fig:8}(a).  In region R3 (plots for $V_1/J=0$, and 0.07 in Fig. \ref{fig:14}), RRIs are either absent or very weak, and if non-zero, their effect becomes visible only at longer times. Therefore, the initial growth of $\mathcal S_A(t)$ is unchanged by the presence of RRIs but they influence the long time behavior ($\bar {\mathcal S}_A$). Also, in region R3, there are no bound states, and RRIs only effectively reduces the strength of exchange couplings that resulting in a lower $\bar {\mathcal S}_A$ compared to that for $V_1=0$, as seen in Figs. \ref{fig:14}(a) and \ref{fig:14}(b).  

In region R2, we have the presence of both bound and scattering states in the excitation dynamics. The bound states enhance the role of two excitation basis states in $\mathcal S_A(t)$, and as a result, we see an increment in the long time value of $\bar {\mathcal S}_A$ as $V_1$ increases, [see Fig.~\ref{fig:14}(b)]. With a further increase of $V_1$ deep into the region R1, the dynamics purely becomes a quantum walk of a bound pair of two excitations. In this case, the basis states having one excitation in each subsystem become redundant, except one state in which one excitation on either side of the border between the subsystems $A$ and $B$. The latter reduces the number of non-zero eigenvalues of the reduced density matrix $\rho_A$, which results not only in the slow growth of $\mathcal S_A(t)$ but also in a smaller long time value.

%%%%%%%%%%

\section{Dissipative Dynamics}
\label{diss}
%%%%%%%%%%
\begin{figure}[hbt]
\centering
\includegraphics[width= 1.\columnwidth]{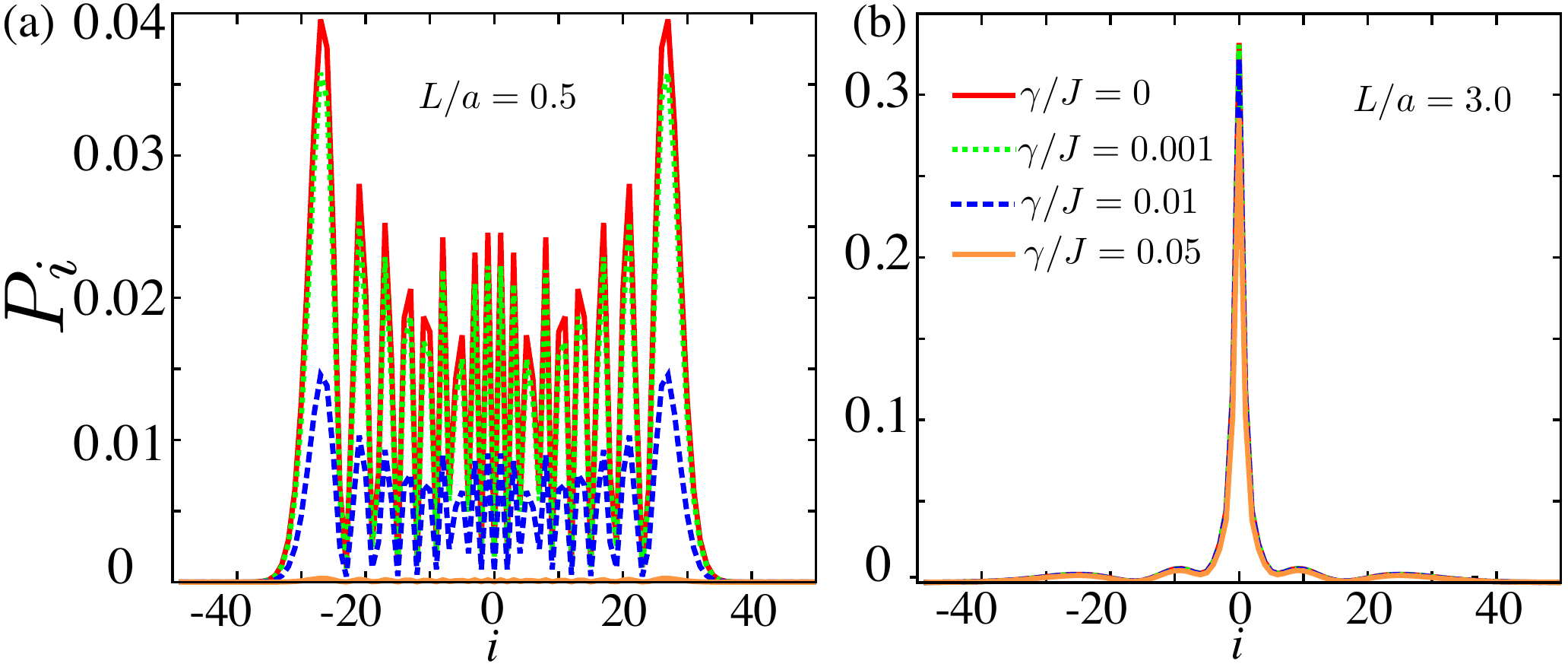}
\caption{\small{ The dissipative single excitation dynamics for (a) $L/a=0.5$, and (b) $L/a=3$ with different values of $\gamma$. The dynamics in (a) resemble that of a CTQW and in (b) shows the quasi-localization behavior. The times at which each of the snapshots are taken (a) $Jt=100$, and (b) $Jt=3$.}}
\label{fig:15} 
\end{figure}
In this section, we briefly outline the effect of dissipation in the dynamics of one and two excitations, as this will be present in any realistic implementation with atoms coupled to a photonic crystal waveguide. We also give estimates of typical parameters to exprimentally observe all the rich Rydberg phenomena we have described.

{\em Including spontaneous emission}.--- The two primary sources of dissipation are the spontaneous emission to the free space from the excited state with a decay rate $\gamma$, and the photon loss from the photonic crystal \cite{dou15}. Since we are interested in the regime where the photonic crystal modes are only  weakly populated, we can safely neglect the latter. In the presence of atomic spontaneous emission, the state of the atomic chain must be described by a density matrix $\rho(t)$, whose dynamics in the Markovian approximation is governed by the master equation, 
\begin{equation}
\dot{\rho} = -\frac{i}{\hbar}[H_t,\rho]+ \mathcal{L}_{\gamma}\rho
\end{equation}
with a Lindblad term of the form,
\begin{equation}
\mathcal{L}_{\gamma}\rho = -\frac{\gamma}{2}\sum_{j}^{}\big( \{ \sigma_{ee}^j,\rho \}-2\sigma_{ge}^j \rho \sigma_{eg}^j  \big).
\end{equation}
In Figs. \ref{fig:15}(a) and \ref{fig:15}(b), we show the single excitation dynamics for different values of $\gamma$ with $L/a=0.5$ and $L/a=3.0$.  As we can see, even though the excitation probability decays drastically, the features of the coherent dynamics are intact for a decay rate of $5\%$ of the hopping strength. For instance, the quasi-localization behavior at large values of $L$ is visible, as shown in Fig. \ref{fig:15}(b). After a sufficiently long time ($t\gg 1/\gamma$), the excitation decays completely, and that halts the dynamics. We verified that CTQW for small $L$ has survived upto a decay rate of $20\%$, i.e., for $\gamma=0.2J$ whereas the quasi-localization at large $L$ is visible even for $\gamma\sim J$ despite having a small excitation probability.

\begin{figure}
\centering
\includegraphics[width= 1.\columnwidth]{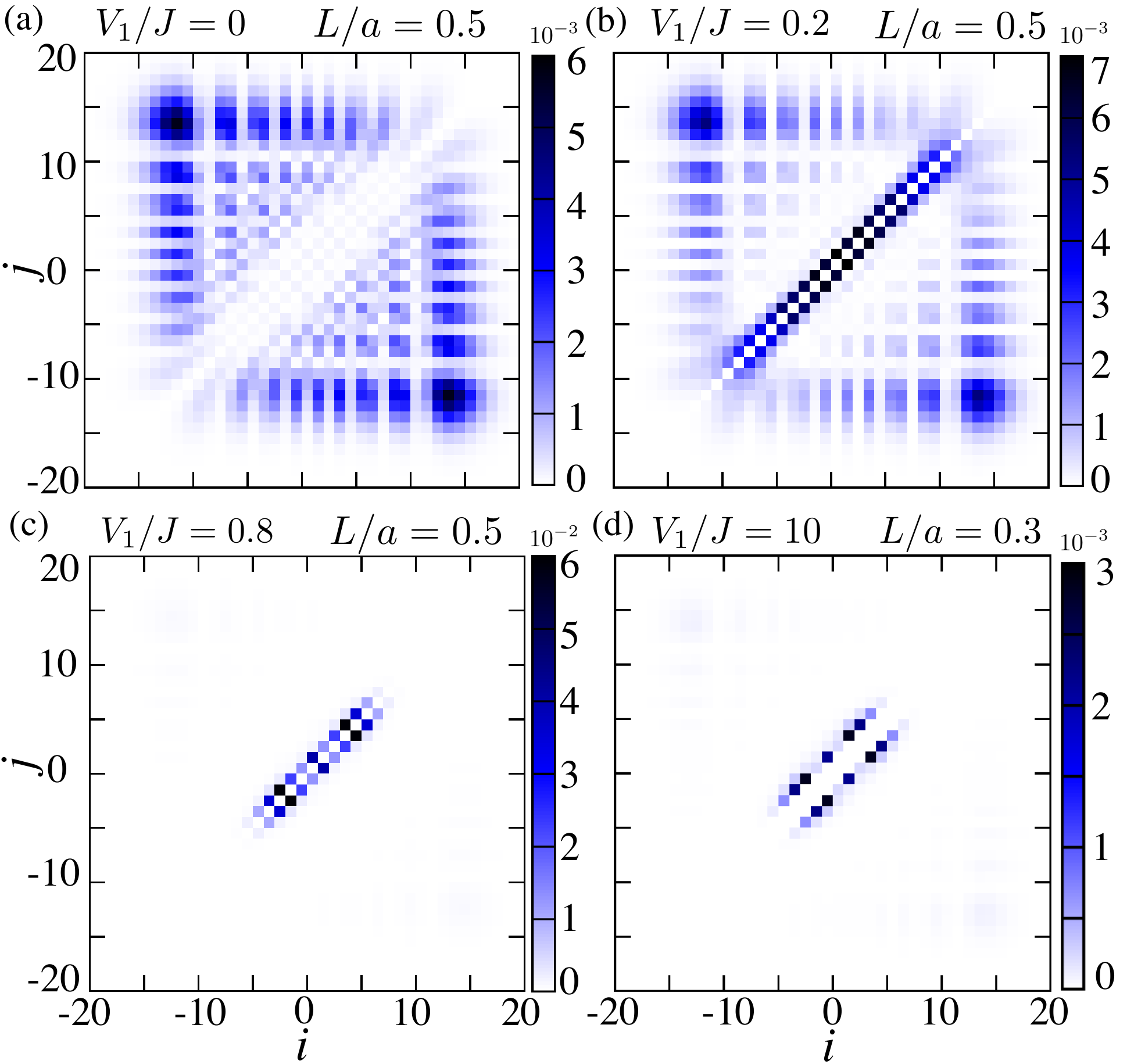}
\caption{\small{ Two-excitation dissipative dynamics for $L/a=0.5$ and (a) $V_1=0$, $V_1/J=0.2$, and $V_1/J=0.8$ with $\gamma/J=0.01$. (a) shows the anit-bunching dynamics, (b) exhibits both anti-bunching and bound state dynamics, and (c)-(d)shows the quantum walk of a bound pair of two excitations. (c) and (d) shows the zeroth and first order bound states, respectively. For (d) $L/a=0.3$ and $V_1=10$. The times at which each of the snapshots are taken (a)-(c) $Jt=50$, and (d) $Jt=200$. }}
\label{fig:16} 
\end{figure}

Similarly, for two excitations, we verified that all features are consistent with the coherent dynamics. In particular, we display the anti-bunching dynamics for $V_1=0$ [Fig. \ref{fig:16}(a)], the presence of both bound states and anti-bunching [Fig. \ref{fig:16}(b)], and the quantum walk of a bound pair of excitations as we increase $V_1$. Figs. \ref{fig:16}(c) and  \ref{fig:16}(d) depict the dynamics of zeroth and first order bound states in Figs. \ref{fig:16}(c) and  \ref{fig:16}(d), respectively. In these  simulations we have taken $\gamma=0.01 J$, and also verified that the features are survived upto a decay rate of $\gamma=0.05 J$. Note that, for the single excitation case, we could afford to have a larger decay rate compared to that for two excitations in order to see all the characteristic features. 

%%%%%%
{\em Experimental parameters}.--- Finally, we propose a realistic set of parameters for an experiment with Rubidium atoms coupled to a PCW. The hopping strength $J$ and the range $L$ of the exchange interactions can be tuned by either varying the properties of PCW (e.g.~band curvature $\alpha$) or of the excited atomic state $|e\rangle$. In particular, the strength of the hopping ($J\propto d^2_{eg}$) and of the RRIs $C_6\propto d^4_{eg}$ can be made significantly large (of several MHz) by choosing a highly excited Rydberg state $|e\rangle$ with a large dipole moment $d_{eg}$. For instance, if we take a Rydberg 45$S_{1/2}$ state of a Rubidium atom, the lifetime is approximately 101 $\mu s$ \cite{bet09} or equivalently $\gamma/J=0.0099$ ($\sim$1\% of $J$) if $J=1$MHz. Therefore, by choosing a significantly high $nS_{1/2}$ Rydberg state, with $n>30$, we strongly reduce the effect of the spontaneous emission. Note that, in order to use Rubidium 
$nS_{1/2}$ Rydberg states, one has to rely on either three-level ladder \cite{lan17} or a lambda scheme \cite{dou15}, which in turn can be used to modify exchange interaction strength $J$ beyond the values set by the PCW structure. Alternatively, one could prepare a n$P_{3/2}$ Rydberg excited state, for which the $C_6$ coefficient exhibits an additional angular dependence $C_6\propto\sin^4\theta$, where $\theta$ is the angle between the quantization axis and radial vector between the two Rydberg excitations. In this way, we can externally control the Rydberg-Rydberg interactions from zero (non-interacting case) to a maximum value by changing the angle $\theta$ for a fixed $n$ \cite{rei07, gla14}.

In the following we comment on how to read out the excitation dynamics in our setup. Since, the atomic resonant frequency lies at the band gap, we cannot map the atomic excitation into the photons of the waveguide modes. One way would be to optically transfer the atomic excitation into a lowest atomic state so that the information can be mapped into guided photons. Other way, which is currently being employed in atomic lattices with Rydberg excitations \cite{bar15}, is to first remove the traps for all atoms, and then use dipole traps to recapture the ground state atoms. The presence of ground state atoms can be detected using Fluorescence imaging. Once the ground state population is estimated, the same for the Rydberg excitations can be calculated by simple subtraction. In this way, a significantly long coherent exchange dynamics of Rydberg excitations via F\"orster resonance, upto 4$\mu$s has been demonstrated in \cite{bar15}.

Currently, there are intense effort to improve nano-photonic device fabrication, and eventually to realize one-dimensional (1D) and two-dimensional (2D) lattices of atoms near the PC waveguides using array of optical tweezers \cite{lua20}. In particular, it has been demonstrated a lattice with 17 tweezer sites with 10$\mu$m separation between the adjacent sites and and a distance of 1 $\mu$m to 10 $\mu$m away from the PCW. The separation between the PCW and atom lattice is sufficiently large compared to the radius of $nS_{1/2}$ Rydberg state with $n=40-100$, which ranges from 108 nm to 740 nm. Thus, one can adjust the separation between atoms and PCW to suppress unwanted effects from stray fields or PCW surfaces \cite{epp14,lan17}.

%%%%%%%%%%

%%%%
%\section{Experimental Considerations}
%Very recently, an interface between a Rydberg excitation and an optical nanofiber has been achieved \cite{kri19}.
%%%%
\section{Summary and Outlook}

In  this  work,  we  analyzed  in  detail  the  dynamics  of  one and two Rydberg excitations in an atomic array coupled to a photonic crystal waveguide, paying particular attention in the interplay  between RRIs  and  long-range hoppings induced by the waveguide. For  an  initially  localized  single  excitation,  we  observed quantum  diffusion,  which  exhibits a  typical  CTQW  for short-range exchange couplings, as well as quasi-localization and tailing behavior for long-range. The dynamics of IPR reveals  that  the  excitation  dynamics  is  super-ballistic  at  the initial  stage,  whereas it  becomes  ballistic at  long  times. In addition, we found that the bipartite entanglement entropy of the system is globally bounded if a single excitation builds the correlations.

In the case of two Rydberg excitations, the RRIs enter into play, and there exists a competition between the scattering and bound states that emerge in the dynamics. The latter leads to non-trivial dynamics for both excitations and the resulting correlations. We observe anti-bunching, co-existence of anti-bunching and bound-state dynamics, and a CTQW of a bound pair of two excitations. We extended the analysis of bipartite entanglement entropy in the presence of two excitations. In contrast to the single excitation case, the entanglement entropy for two excitations is not globally bounded as it depends on the system parameters. Moreover, we analyzed the two-point correlation function for two excitations, which reveals a light-cone behavior even for sufficiently long-range exchange couplings. 

We also discussed the dissipative dynamics, including imperfections due to the spontaneous emission from the Rydberg state into  photons  outside  the  waveguide. We showed that the features of the coherent dynamics for one and two Rydberg extiations are intact up to reasonable decay rates of 5$\%$ of the hopping rate. The complex dynamics of the Rydberg excitations and their quantum correlations when propagating through the waveguide could be measured via photo-detection or multi-photon homodyne measurement techniques at the photonic outputs of the waveguide \cite{leg03,sil10,can14,ram17,tur19}. 

The excitation transport has been previously studied in Rydberg gases but mediated by dipole-dipole interactions involving multiple Rydberg states (via F\"orster resonances) \cite{gun13,sch15,sche15,bar15} and the environment also played a crucial role in the transport properties. Therefore, one new direction would be to probe the role of the environment in the quantum diffusion of Rydberg excitations in a PC setup. Besides, if multiple Rydberg states are involved, the competition between dipole-dipole and cavity mediated exchange interactions may lead to non-trivial scenarios. Further, if we allowed to overlap the Rydberg orbitals of two excitations, the strong coulomb interactions between the delocalized electrons may lead to novel correlated phases, assuming the Penning ionization rates can be significantly suppressed \cite{mic20}. Our analysis on bound states of two excitations can shed new  light  into  the  mechanisms  for  stabilizing  bound  states by repulsive interactions,  using a completely different setup compared to the traditional Hubbard models. In addition, our approach  can  be  extended  to  three  or  more  excitations, where  the emergence  of three-body  bound states would be very relevant in the context of Efimov physics \cite{nai17}. 

\section{Acknowledgments}
R. N. acknowledges UKIERI- UGC Thematic Partnership No. IND/CONT/G/16-17/73 UKIERI-UGC project. T. R.  acknowledges funding from the EU Horizon 2020 programme under the Marie Sklodowska-Curie grant agreement No 798397.
%%%%%%%%%%%%%%%%%%%%%%%%%%
\appendix
\section{Single excitation: Dispersion relation, width and kurtosis of the probability distribution}	
\label{a1}
The Hamiltonian of the system is given by,
	\begin{equation}
	\hat{H}_{ex} = J \sum_{j\ne l}^{} e^{-|x_j-x_l|/L} \hat \sigma^{j}_{eg}\hat \sigma^{l}_{ge}
	\end{equation} 
and the corresponding Schr\"odinger equation is,
    \begin{equation}
	i \hbar \frac{d}{dt} \ket{\psi} = \hat{H}\ket{\psi}.
	\end{equation}  
Taking $\ket{\psi(t)} = \sum_{i} c_i(t)\ket{n}$ where $\{|i\rangle\equiv |...g^{i-1} e^i g^{i+1}... \rangle \}$ provides the single excitation basis, we can rewrite the Schr\"odinger equation as,
          \begin{equation}
          \label{ae1}
	 i\frac{dc_j}{dt} -J \sum_{l\neq j}^{N} e^{-|j-l|a/L}c_l =0.
	\end{equation}
with $\hbar=1$. To get the dispersion relation for the plane waves we take the ansatz of the form $c_j(t) = Ae^{i(jka+\omega_k t)}$. Substituting this into Eq. (\ref{ae1}) we get the dispersion relation:
	 \begin{equation}
	\omega_k = -2J\sum_{d=1}^{\infty}e^{-da/L}\cos(kad).
	\end{equation}  
	After the summation,	
 \begin{equation}
 \label{wk}
\omega_k = \frac{J(\cos ka-e^{-a/L})}{(\cosh (a/L)-\cos ka)}.
\end{equation}
Using the dispersion relation, below we obtain the width and the Kurtosis of the probability distribution. 

The second moment of the probability distribution is given by $\langle z^2 \rangle=a^2\sum_{n=-\infty}^{\infty} n^2 p_n(t)$ where $p_n(t) = |u_n(t)|^2$ with $u_n(t) = \frac{a}{2\pi} \int_{BZ} \exp(i[kna-\omega_k t])dk$. With a little algebra we rewrite it as,
\begin{equation}
\langle z^2\rangle =\bigg[ \frac{a}{2\pi} \int_{BZ} \bigg(\frac{d\omega_k}{dk}\bigg)^2 dk\bigg] t^2.
\end{equation}
Using the dispersion in Eq. (\ref{wk}) and upon integrating we get,
\begin{equation}
\langle z^2 \rangle = \frac{J^2a^2}{2} \frac{\coth(a/L)}{\sinh(a/L)^2} t^2,
\label{z2}
\end{equation}
which then gives us 
\begin{equation}
\alpha(L)=aJ\sqrt{\frac{\coth(a/L)}{2\sinh^2(a/L)}}.
\end{equation}

Similarly, the excess kurtosis $\kappa$ of the distribution is given by,
\begin{equation}
\kappa = \frac{\langle z^4 \rangle}{(\langle z^2 \rangle)^2}-3
\label{kurt}
\end{equation}
The fourth moment of the distribution $\langle z^4 \rangle = a^4\sum_{n=-\infty}^{\infty} n^4 p_n(t)$ in terms of dispersion is given by,
\begin{equation}
\langle z^4 \rangle = \frac{a}{2\pi}  \int_{BZ}  dk  \bigg[ \bigg(\frac{d^2\epsilon_k}{dk^2}\bigg)^2t^2+\bigg(\frac{d\epsilon_k}{dk}\bigg)^4t^4\bigg]
\end{equation}
 Finally we get,
 \begin{eqnarray}
 \langle z^4 \rangle &=& J^2a^4\frac{\cosh(a/L)\left[1+\frac{1}{2}\cosh^2(a/L)\right]}{\sinh^5(a/L)} t^2 \nonumber \\
 &&+\frac{3J^4a^4}{16} \frac{\cosh(a/L)\left[1+2\cosh^2(a/L)\right]}{\sinh^7(a/L)}t^4
 \label{z4}
 \end{eqnarray}
Substituing equation \ref{z4} and \ref{z2} in \ref{kurt} and after simplification we obtain,
\begin{equation}
\kappa(t) = 4\left[1+\frac{1}{2}\cosh^2(a/L)\right]\frac{1}{(Jt)^2} +\frac{3}{2}  \frac{[1+2\cosh^2(a/L)]}{\sinh(2a/L)}-3.
\end{equation}
The asymptotic value ($t\to\infty$) of kurtosis is given by
\begin{equation}
\kappa_a = \lim\limits_{t\rightarrow \infty} \kappa(t) =  \frac{3}{2}  \frac{[1+2\cosh^2(a/L)]}{\sinh(2a/L)}-3
\end{equation}
Further in the large $L$ limit where $L/a \gg 1$, we can use $\cosh(a/L) \approx 1 $ and $\sinh (a/L) \approx a/L$ to obtain $\kappa_a \approx 9L/(4a)-3$.
%%%%%%%%%%%%%%%%%%%

\section{Two non-interacting excitations ($C_6=0$): Dynamics for different $d_0$ and $L$.}	
\label{a2}

%%%%%
\begin{figure}[hbt]
\centering
\includegraphics[width= 1.\columnwidth]{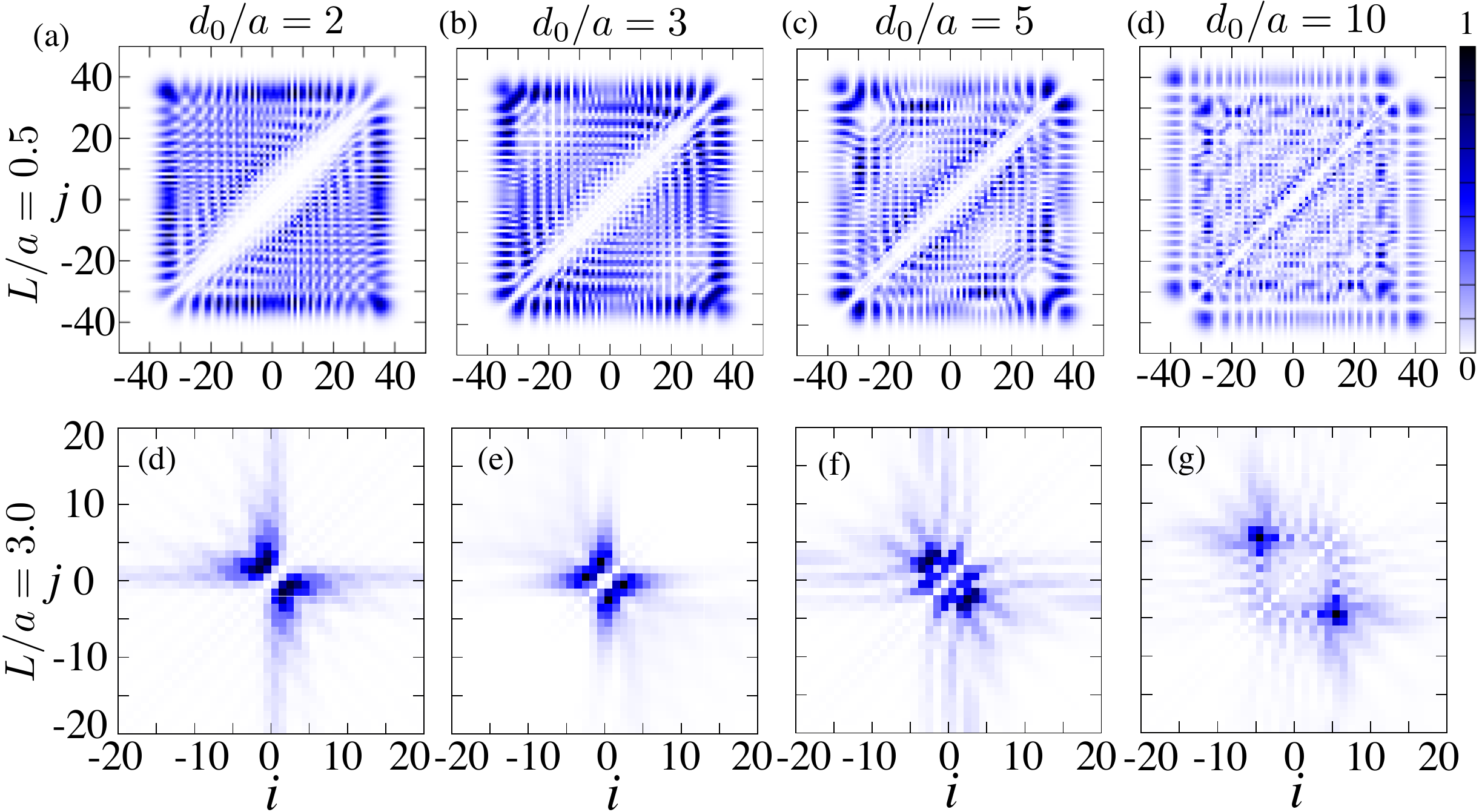}
\caption{\small{The two-excitation LTPD $\Gamma_{ij}$ for $C_6=0$ with different values of $d_0$ and $L$. Along the row $L$ is fixed, and along columns $d_0$ is fixed. The times at which each of the snapshots are taken (a)-(d) $Jt=125$, and (e)-(f) $Jt=3$.}}
\label{fig:a1} 
\end{figure}
%%%%%%%
In Fig. \ref{fig:a1}, we show two-excitation LTPD $\Gamma_{ij}$  for $C_6=0$ with different values of $d_0$ and $L$. In contrast to Fig. \ref{fig:5}(a) (main text), the spatial anti-bunching at small $L$ becomes less prominent when $d_0>1$, due to the interference effects (along the first row in Fig. \ref{fig:a1}). In contrast, the quasi-localization of the single excitations at large $L$ becomes more prominent as $d_0$ gets larger (the second row in Fig. \ref{fig:a1}).

%%%%%%%%%%%
\bibliographystyle{apsrev4-1}
\bibliography{liball.bib}
\end{document}